\documentclass[aps,pre,showpacs,showkeys,onecolumn,10pt]{revtex4-1}
\usepackage[dvips]{graphicx}
\usepackage[dvips]{color}
\usepackage{hyperref,breakurl,amsmath,amssymb,pst-node,eepic}
\usepackage{bbm}
\usepackage{tikz}
\usepackage{verbatim}
\usepackage{bm}
\usetikzlibrary{shapes}
\usepackage{times}

\definecolor{light-gray}{gray}{0.75}
\definecolor{vlight-gray}{gray}{0.9}

\begin{document}

\title{Eigenvector statistics of the product of Ginibre matrices}

\author{Zdzis{\l}aw Burda}
\email{zdzislaw.burda@agh.edu.pl}
\affiliation{AGH University of Science and Technology,  
Faculty of Physics and Applied Computer Science, 
al. Mickiewicza 30, 
30-059 Krak{\'o}w, Poland}
\author{Bart{\l}omiej J. Spisak}
\email{bjs@agh.edu.pl }
\affiliation{AGH University of Science and Technology,  
Faculty of Physics and Applied Computer Science,
al. Mickiewicza 30, 
30-059 Krak{\'o}w, Poland}
\author{Pierpaolo Vivo}
\email{pierpaolo.vivo@kcl.ac.uk}
\affiliation{Department of Mathematics, 
King's College London,
Strand WC2R 2LS, 
London, U.K.}

\begin{abstract}
We develop a method to calculate left-right eigenvector correlations of the 
product of $m$ independent $N\times N$ complex Ginibre matrices.
For illustration, we present explicit analytical results for the vector overlap 
for a couple of examples for small $m$ and $N$. 
We conjecture that the integrated overlap between left and right eigenvectors 
is given by the formula $O = 1 + (m/2)(N-1)$ 
and support this conjecture by analytical and numerical calculations. 
We derive an analytical expression for the limiting correlation density as 
$N\rightarrow \infty$ for the product of Ginibre matrices as well as for 
the product of elliptic matrices. 
In the latter case, we find that the correlation function is independent of the 
eccentricities of the elliptic laws.

\keywords{random matrix theory, non-hermitian, planar diagram enumeration}

\end{abstract}

\maketitle
\section{Introduction}

Products of random matrices have continuously attracted attention since the 
sixties~\cite{fk,o,bl,ckn,cpv}. 
They are of relevance in many fields of mathematics, physics and engineering 
including dynamical systems~\cite{o,r}, disordered systems~\cite{dh,cln,l}, 
statistical mechanics~\cite{gjjnw}, quantum mechanics~\cite{gmw}, quantum 
transport and mesoscopic systems~\cite{a,b}, hidden Markov models~\cite{em}, 
image processing~\cite{jljn}, quantum chromodynamics~\cite{lnw}, wireless 
telecommunication~\cite{m,cd}, quantitative finance~\cite{pbl,blmp,bjjnpz} and 
many others~\cite{abf}. 
Recently, an enormous progress has been made in the understanding of  
macroscopic~\cite{v,vdn,jw,k,bjlns,bjw,agt,gt,bjn,rs,pz,bns,bls,bcs,bms,gnt, 
rrsv,bstv,gkt,bss,s,bs} 
and microscopic~\cite{ab,akw,as,aik,i,l2,abkn,ais,f,f2,ik,ks,lw,n,ckw,hjl,k2, 
ai,kks,i2,arrs,kk}
statistics of eigenvalues and singular values as well as of Lyapunov spectra 
for products of random matrices~\cite{n2,in,bzq,k3,p,v2,f3,k4,abk,i3,f4}.
In contrast, not much has been learned about the eigenvector statistics of the 
products of random matrices so far. 
In this paper, we address this problem by considering a correlation function 
for eigenvectors of the product of Ginibre matrices. 
More precisely, we study the overlap between left and right eigenvectors for 
finite $N$ and for $N\rightarrow \infty$. 
In the first part of the paper, we adapt ideas developed in~\cite{cm,mc} to the 
product of random matrices by using the generalized Schur 
decomposition~\cite{ab} for finite $N$, while in the second part we combine the 
generalized Green function method~\cite{rj,jnpz,jnpwz,jnnpz} with 
linearization (subordination)~\cite{gjjnw,bms,bjw} to derive the limiting law 
for the overlap for $N\rightarrow \infty$.  

\section{Definitions}

Consider a diagonalizable matrix $X$ over the field of complex numbers.
Let $\{\Lambda_\alpha\}$ be the eigenvalues of $X$. The corresponding left 
eigenvectors $\left\langle L_\alpha\right|$ and right eigenvectors 
$\left| R_\alpha\right\rangle$ satisfy the relations
\begin{equation}
X\left|R_\alpha \right\rangle 
= 
\Lambda_\alpha \left|R_\alpha\right\rangle ,
\quad \left\langle L_\alpha \right| 
X 
= 
\left\langle L_\alpha \right| \Lambda_\alpha  \ .
\end{equation}
Note that the Hermitian conjugate of the second equation has the form: 
$X^\dagger \left|L_\alpha \right\rangle 
= 
\bar{\Lambda}_\alpha \left|L_\alpha \right\rangle$, 
where the symbol 'bar' denotes the complex conjugation of $\Lambda_\alpha$.
The eigenvectors fulfill the bi-orthogonality and closure relations in the form
\begin{equation}
\label{bort}
\left\langle L_\alpha | R_\beta \right\rangle 
= 
\delta_{\alpha\beta}\ , \quad
\sum_\alpha \left| L_\alpha \right\rangle \left\langle R_\alpha \right| 
= 
1\ .
\end{equation}
The two relations are invariant with respect to the scale transformation
\begin{equation}
\label{st}
\left| R_\alpha \right\rangle \rightarrow c_\alpha \left| R_\alpha \right\rangle  ,
\quad 
\left\langle  L_\alpha \right| \rightarrow 
\left\langle L_\alpha \right| c_\alpha^{-1}\ ,
\end{equation}
with arbitrary non-zero coefficients $c_\alpha$'s. 
According to Refs.~\cite{cm,mc}, an overlap of the left and right eigenvectors 
is defined in the following way
\begin{equation} 
O_{\alpha\beta} 
= 
\left\langle L_\alpha | L_\beta \right\rangle 
\left\langle R_\beta | R_\alpha \right\rangle\ .
\label{leftrightoverlap}
\end{equation}
By construction, the quantity $O_{\alpha\beta}$ is invariant with respect to 
the scale transformation given by Eq.~\eqref{st} and consequently does not 
depend on the vector normalizations.

If $X$ is a random matrix, one defines averages over the ensemble
\begin{equation}
\langle O_{\alpha\beta} \rangle 
= 
\int d\mu(X) O_{\alpha\beta}\ ,
\end{equation}
where $d\mu(X)$ is the probability measure for the random matrix in question.
The dependence of $O_{\alpha\beta}$ on $X$ is suppressed in the notation.
We use this notation throughout the paper also for other observables
that depend on random matrices. 
The global diagonal overlap averaged over the ensemble is given by
\begin{equation}
O 
= 
\left\langle \frac{1}{N} \sum_{\alpha=1}^N O_{\alpha\alpha} \right\rangle,
\end{equation}
while the global off-diagonal one is expressed by the formula
\begin{equation}
\label{Oo}
O_{off} = 
\left\langle \frac{2}{N(N-1)} \sum_{\alpha<\beta} O_{\alpha\beta} 
\right\rangle\ .
\end{equation}
We are interested here in unitarily invariant random matrices for which 
the probability measure is invariant with respect to the similarity transformation 
$X \rightarrow U X U^{-1}$, where $U$ is a unitary matrix. 
In particular, this invariance implies that 
$\langle O_{\alpha\alpha} \rangle = \langle O_{11} \rangle$
and 
$\langle O_{\alpha\beta} \rangle = \langle O_{12} \rangle$ 
for any $\alpha$ and $\beta$. 
It follows that
\begin{equation}
\label{ddent}
O
= 
\left\langle O_{11} \right\rangle\ , \quad
O_{off} 
= 
\left\langle O_{12} \right\rangle\ .
\end{equation}
We can also define the local diagonal overlap density by the formula
\begin{equation}
O(z) 
= 
\left\langle \frac{1}{N} \sum_{\alpha=1}^N O_{\alpha\alpha} 
\delta\left(z-\Lambda_\alpha\right)\right\rangle 
= 
\left\langle O_{11} \delta\left(z-\Lambda_1\right)\right\rangle,
\label{localO}
\end{equation}
and the off-diagonal one by
\begin{equation}
\begin{split}
O_{off}(z,w) & 
= 
\left\langle \frac{2}{N(N-1)} \sum_{\alpha<\beta} O_{\alpha\beta}
\delta\left(z-\Lambda_\alpha\right) \delta\left(w-\Lambda_\beta\right)  
\right\rangle \\
& 
= 
\left\langle O_{12} \delta\left(z-\Lambda_1\right)
\delta\left(w-\Lambda_2\right) \right\rangle\ .
\end{split}
\end{equation}
The symbol $\delta(z)$ denotes the Dirac delta function
on the complex plane.
Clearly, the diagonal global overlap is equal to the integrated overlap density 
given by Eq.~\eqref{localO}, i.e.
\begin{equation}
O 
= 
\int d^2 z\ O(z)\ .
\end{equation}

\section{Product of Ginibre matrices}

Consider the product 
\begin{equation}
X = X_1 X_2 \cdots X_m
\label{product}
\end{equation}
of $m$ independent identically distributed $N\times N$ Ginibre random 
matrices~\cite{g} with complex entries. 
The probability measure factorizes and can be written as a product of measures 
for individual Ginibre matrices 
\begin{equation}
d\mu(X) \equiv d\mu(X_1,X_2,\ldots,X_m) 
= 
d\mu(X_1) d\mu(X_2) \cdots d\mu(X_m)\ ,
\end{equation}
each of which is given by
\begin{equation}
d\mu(X_i) 
=
 \left(\pi\sigma^2\right)^{-N^2} e^{-\frac{1}{\sigma^2} 
\mathrm{Tr} X_iX_i^\dagger} DX_i\ ,
\label{individualG}
\end{equation} 
where $\sigma$ is a scale parameter, and
$DX_i 
= 
\prod_{\alpha\beta} 
d \mathrm{Re} X_{i,\alpha\beta} d \mathrm{Im} X_{i,\alpha\beta}$. 
According to Eq.~\eqref{localO}, the local diagonal overlap density can be 
calculated with respect to the measure $d\mu(X)$ in the following way
\begin{equation}
O(z) 
= 
\int  d\mu(X) O_{11} \delta(z-\Lambda_1)\ ,
\label{od}
\end{equation}
where $\Lambda_\alpha$'s correspond to the eigenvalues of the product $X$
\eqref{product}. An analogous formula holds for the off-diagonal density.
In the calculations we set $\sigma=1$. 
One can easily transform the result to other values of $\sigma$ 
(\ref{individualG}) by using the formula
\begin{equation}
O_{\sigma}(z) 
=
\frac{1}{\sigma^{2m}} O_{\sigma=1}\left(\frac{z}{\sigma^{m}}\right)\ ,
\end{equation}
which merely corresponds to the scale transformation of all Ginibre matrices 
$X_i \longrightarrow \sigma X_i$ in the product~(\ref{product}).
Later, when discussing the limiting laws for $N\rightarrow \infty$ we will 
choose $\sigma = N^{-1/2}$. 
This choice of the scale parameter $\sigma$ will ensure the existence of the 
limiting eigenvalue density on a compact support being the unit disk in the 
complex plane. 

\section{Calculations of the overlap for finite {\boldmath $N$}}

In order to calculate the global left-right vector overlap, defined by 
Eq.~\eqref{leftrightoverlap}, for the product of Ginibre 
matrices~\eqref{product}, we will change the parametrization of the matrices 
$X_i$'s using the generalized Schur decomposition~\cite{ab}
\begin{equation} 
X_i 
= 
U_{i-1} \tau_i U_i^\dagger\ , 
\label{gschur}
\end{equation}
for $i=1,\ldots,m$, where $U_i$ are unitary matrices from the unitary group 
$U(N)$, and $\tau_i$ are upper triangular matrices of size $N\times N$.
We use a cyclic indexing $U_i\equiv U_{m+i}$, in particular $U_0\equiv U_m$. 
Sometimes it is convenient to express each $\tau_i$ as a sum of a diagonal 
matrix $\lambda_i$ and a strictly upper triangular one $t_i$, namely
\begin{equation}
\tau_i=\lambda_i + t_i = \left(\begin{array}{llllll} 
\lambda_{i,1} & t_{i,12}     & t_{i,13} & \ldots & t_{i,1N} \\
0             & \lambda_{i,2} & t_{i,23} & \ldots & t_{i,2N} \\
          &   &              & \ddots        &  \\
0             &     0        & 0  & \lambda_{i,N-1}  & t_{i,N-1N} \\
0             &     0        & 0  & \ldots  & \lambda_{i,N} \end{array} \right)\ .
\end{equation}
In this representation, the product $X$ is unitarily equivalent to a matrix 
$\mathcal{T}$, that is $X=U_m \mathcal{T} U^\dagger_m$,  where
\begin{equation}
\mathcal{T} 
=
\tau_1\tau_2 \cdots \tau_m\ .
\end{equation}
The matrix $\mathcal{T}$ has also an upper triangular form
\begin{equation}
\label{Tau}
\mathcal{T}=\Lambda + T = \left(\begin{array}{llllll} 
\Lambda_{1} & T_{12}     & T_{13} & \ldots & T_{1N} \\
0             & \Lambda_{2} & T_{23} & \ldots & T_{2N} \\
          &   &              & \ddots        &  \\
0             &     0        & 0  & \Lambda_{N-1}  & T_{N-1N} \\
0             &     0        & 0  & \ldots  & \Lambda_{N} \end{array} \right)\ .
\end{equation}
The diagonal elements of $\mathcal{T}$ are given by
\begin{equation}
\mathcal{T}_{\alpha} 
\equiv \Lambda_\alpha  
=
\lambda_{1,\alpha}\lambda_{2,\alpha} \cdots \lambda_{m,\alpha}\ ,
\end{equation}
and the off-diagonal ones by
\begin{equation}
\mathcal{T}_{\alpha\nu} 
= 
\sum_{\alpha\le\beta \le\ldots\le\nu} 
\tau_{1,\alpha\beta} \tau_{2,\beta\gamma}\cdots\tau_{m,\mu\nu}\ .
\end{equation}
Any instance of $\tau_{i,\alpha\alpha}$ with two identical Greek indices can 
be replaced by $\lambda_{i,\alpha}$ and of $\tau_{i,\alpha\beta}$ with
two different Greek indices by $t_{i,\alpha\beta}$ in the last formula.
One can also express the integration measure in terms of $U$'s, $\lambda$'s
and $t$'s. 
Since one is interested in invariant observables, the $U$'s can be integrated 
out. For the scale parameter $\sigma=1$ one gets~\cite{ab}
\begin{equation}
d\mu(\lambda,t) 
=
Z^{-1} \left|\Delta\left(\bm\Lambda\right)\right|^2
\prod_{i,\alpha} 
e^{-|\lambda_{i,\alpha}|^2} d^2\lambda_{i,\alpha} 
\prod_{j,\beta<\gamma} 
\frac{1}{\pi} e^{-|t_{j,\beta\gamma}|^2}
d^2 t_{j,\beta\gamma}\ ,
\label{mlt}
\end{equation}
where the normalization factor $Z$ is given by the formula
\begin{equation} 
Z = N! [\pi^N 1!2!\cdots (N-1)!]^m \ ,
\label{Z}
\end{equation}
and the Vandermonde determinant $\Delta\left(\bm\Lambda\right)$ for the product 
$X = X_1 X_2 \cdots X_m$ has the form 
\begin{equation}
\Delta\left(\bm\Lambda\right) 
= 
\prod_{\alpha<\beta} 
\left(\lambda_{1,\alpha}\lambda_{2,\alpha}\cdots\lambda_{m,\alpha}-
\lambda_{1,\beta}\lambda_{2,\beta}\cdots\lambda_{m,\beta} \right) 
=
\prod_{\alpha<\beta} 
\left(\Lambda_{\alpha} - \Lambda_{\beta} \right) \ . 
\end{equation}
The square of the determinant in Eq.~\eqref{mlt} comes from the Jacobian of 
the transformation~\eqref{gschur}. 

The next step is to express the observables in terms of $t's$ and $\lambda's$. 
For example, to calculate the diagonal overlap density [cf. Eq.~\eqref{od}], 
we have to find $O_{11}=O_{11}(t,\lambda)$ and to integrate over $t$'s and 
$\lambda$'s with the Dirac delta constraint
\begin{equation}
O(z) 
=
\int d\mu(\lambda,t) O_{11}(t,\lambda) \delta(z-\Lambda_1)\ ,
\label{over_den}
\end{equation}
while for the global overlap 
$O=\int d\mu(\lambda,t) O_{11}(t,\lambda)$.
The measure $d\mu(\lambda,t)$ (\ref{mlt}) factorizes 
$d\mu(\lambda,t)=d\mu(\lambda)d\mu(t)$.
One can first integrate over $t$'s. 
This is a Gaussian integral and can be easily performed. 
After this integration, only the dependence on $\lambda$'s is left 
\begin{equation}
O_{11}(\lambda) 
=
\int d\mu(t) O_{11}(t,\lambda)\ ,
\label{int_over_ts}
\end{equation} 
where $d\mu (t)$ is a normalized Gaussian measure equal to 
the $t$-dependent piece of $d\mu(\lambda,t)$ \eqref{mlt}.
The last step is to integrate over $\lambda$'s with the measure given by 
Eq.~\eqref{mlt}
\begin{equation}
O(z) 
= 
Z^{-1} \int d\mu(\lambda) \left|\Delta\left(\bm\Lambda\right)\right|^2
e^{-\sum_{i,\alpha} |\lambda_{i,\alpha}|^2} O_{11}(\lambda) \delta(z-\Lambda_1)\ ,
\label{Oz}
\end{equation}
where as before $\Lambda_\alpha$'s stand for 
$\Lambda_\alpha
=
\lambda_{1,\alpha} \lambda_{2,\alpha} \cdots \lambda_{m,\alpha}$.  
We will do this below. 
First we have to find the function $O_{11}(t,\lambda)$.
This can be done as follows.
We choose the basis in which the product matrix $X$ is equal to $\mathcal{T}$. 
Such a basis exists since the two matrices are unitarily equivalent.
In this basis, the first right eigenvector $|R_1\rangle$ is represented as a column 
vector with '$1$' in the position $1$ and zeros elsewhere: 
$|R_1\rangle=(1,0,0,\ldots)^T$. 
The vector is written here as transpose of a row vector to save space. 
Denote the elements of the first left eigenvector 
$\langle L_1|=(B_1,B_2,\ldots)$.
The eigenvalue equation $\langle L_1| \mathcal{T} = \langle L_1| \Lambda_1$
leads to the following recursion relation for $B_\beta$'s~\cite{cm,mc}
\begin{equation}
B_\beta 
=
\frac{1}{\Lambda_1 - \Lambda_\beta} \sum_{\alpha=1}^{\beta-1}
B_\alpha T_{\alpha\beta}\ .
\end{equation}
The recursion is initiated by $B_1 = 1$ as follows from the bi-orthogonality
relation~\eqref{bort}. One finds
\begin{equation}
\begin{split}
B_1 &=1  , \\
B_2 &= \frac{T_{12}}{\Lambda_1-\Lambda_2}  , \\
B_3 &= \frac{T_{13}}{\Lambda_1-\Lambda_3} + 
\frac{T_{12}T_{23}}{(\Lambda_1-\Lambda_2)(\Lambda_1-\Lambda_3)}  , \\ 
B_4 &= \frac{T_{14}}{\Lambda_1-\Lambda_4} +
\frac{T_{12}T_{24}}{(\Lambda_1-\Lambda_2)(\Lambda_1-\Lambda_4)} + 
\frac{T_{13}T_{34}}{(\Lambda_1-\Lambda_3)(\Lambda_1-\Lambda_4)} + \\
& + \frac{T_{12}T_{23}T_{34}}
{(\Lambda_1-\Lambda_2)(\Lambda_1-\Lambda_3)(\Lambda_1-\Lambda_4)} , \quad
\mathrm{etc.}
\end{split}
\label{Bs}
\end{equation}
The element $O_{11}$ of the overlap matrix is related to $B$'s as 
\begin{equation}
O_{11} 
= 
\sum_{\alpha=1}^N |B_\alpha|^2\ ,  
\label{OB}
\end{equation}
and $B$'s depend on $t$'s and $\lambda$'s through $T$'s and $\Lambda$'s.
Combining Eqs.~\eqref{Bs},\eqref{OB} with Eq.~\eqref{over_den} we obtain
an explicit form of the integral over $t$'s and $\lambda$'s which can be done.   
We will give a couple of examples below.

\section{Examples}

\begin{figure}
\includegraphics[width=15cm]{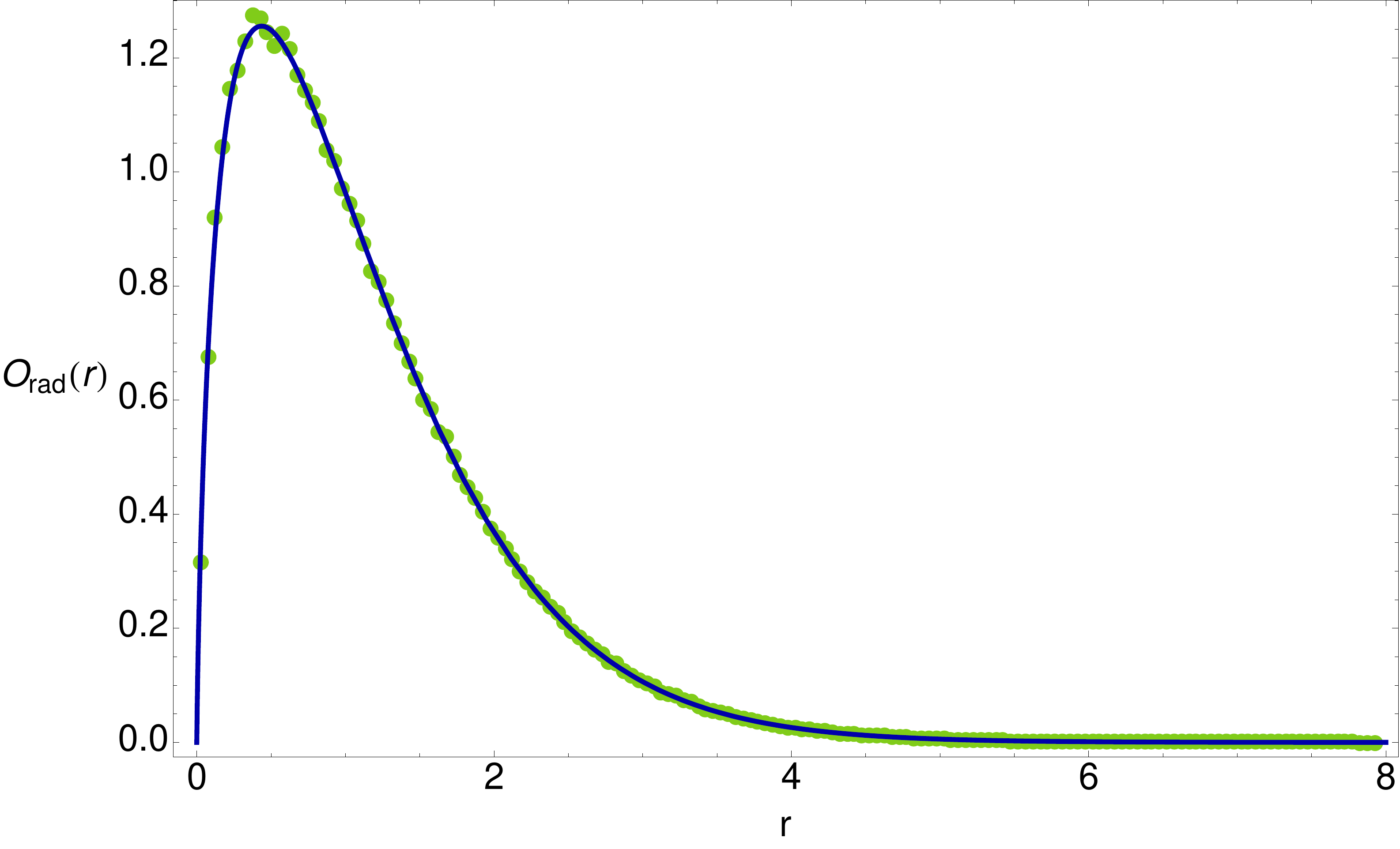}
\caption{
Overlap density for $N=2$ and $m=2$: 
theoretical prediction given by Eq.~(\ref{Or22}) (solid line) and numerical 
histogram (points) generated in Monte Carlo simulations of $10^6$ products of 
two $2\times 2$ Ginibre matrices.
}
\label{fig:product_N2_m2}
\end{figure}

\begin{figure}
\includegraphics[width=15cm]{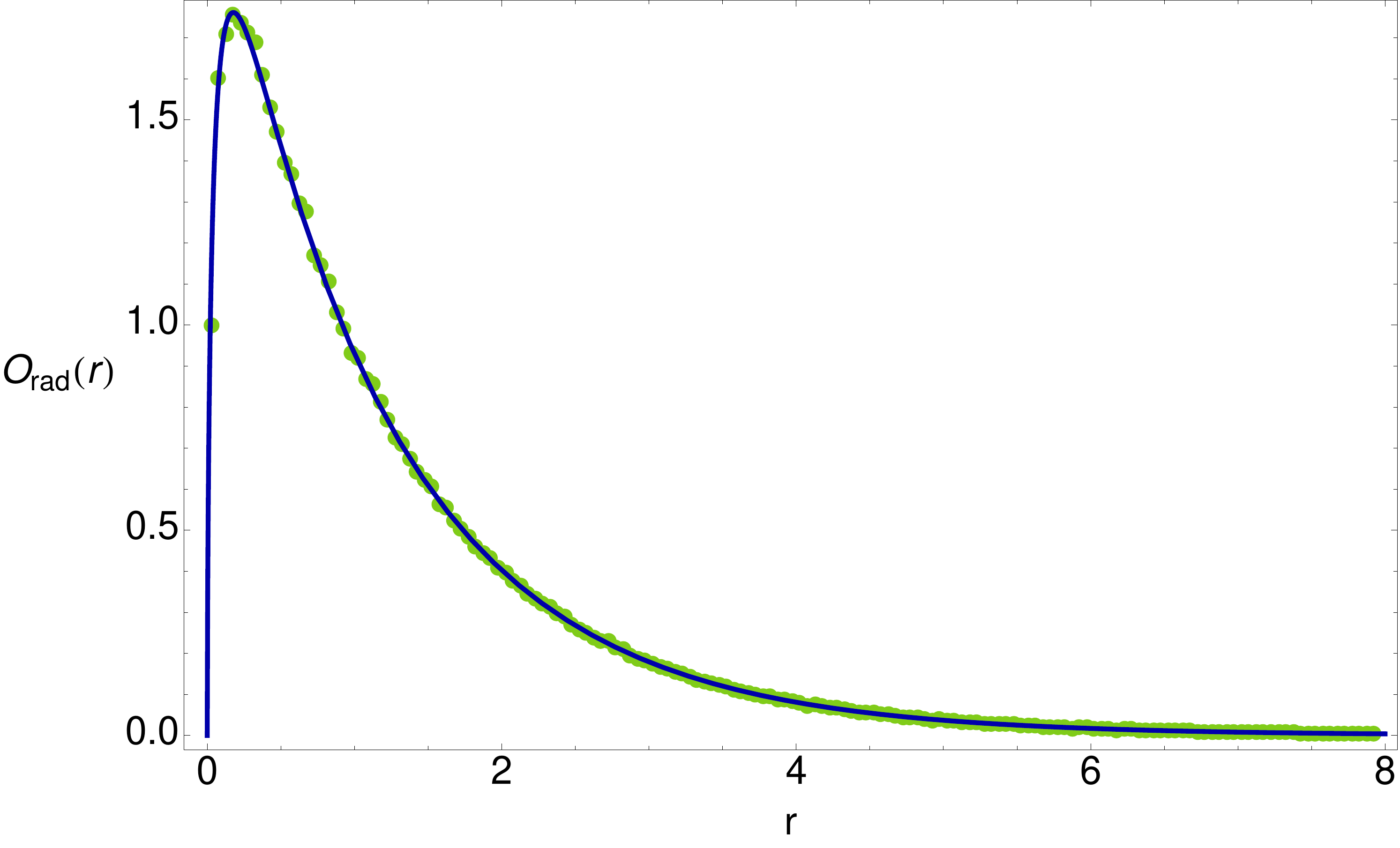}
\caption{
Overlap density for $N=2$ and $m=3$: 
theoretical prediction given by Eq.~(\ref{Or23}) (solid line) and numerical 
histogram (points) generated in Monte Carlo simulations of $10^6$ products of 
three $2\times 2$ Ginibre matrices.
}
\label{fig:product_N2_m3}
\end{figure}

\begin{figure}
\includegraphics[width=15cm]{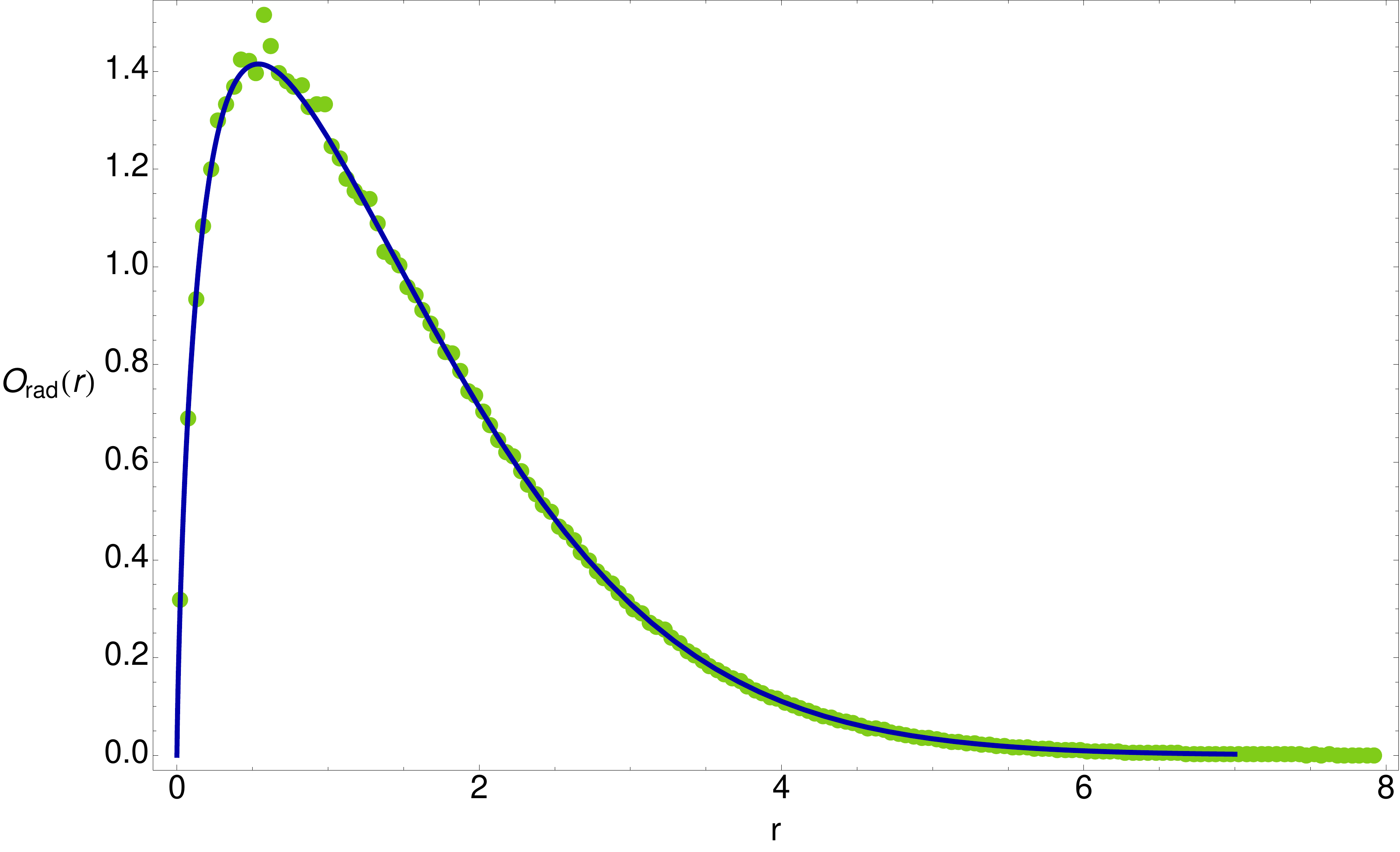}
\caption{
Overlap density for $N=3$ and $m=2$: 
theoretical prediction given by Eq.~(\ref{Or32}) (solid line) and numerical 
histogram (points) generated in Monte Carlo simulations of $10^6$ products of 
two $3\times 3$ Ginibre matrices.
}
\label{fig:product_N3_m2}
\end{figure}

Let us first illustrate the calculations for $N=2$, $m=2$ and $\sigma=1$ - 
that is for the product of two $2\times 2$ Ginibre matrices. 
Firstly, we express $T_{12}$ in terms of $t$'s and $\lambda$'s as follows
\begin{equation}
\mathcal{T} 
= 
\left(\begin{array}{rr} 
\lambda_{1,1} & t_{1,12} \\ 
0 & \lambda_{1,2} \end{array}\right)
\left(\begin{array}{rr} 
\lambda_{2,1} & t_{2,12} \\ 
0 & \lambda_{2,2} \end{array}\right) 
=
\left(\begin{array}{rr} 
\Lambda_1 & T_{12} \\ 
0 & \Lambda_2 \end{array}\right)\ .
\end{equation}
This gives $T_{12} = \lambda_{1,1} t_{2,12} + t_{1,12} \lambda_{2,2}$ and
$\Lambda_\alpha = \lambda_{1,\alpha}\lambda_{2,\alpha}$ for $\alpha=1,2$.
Thus we have
\begin{equation}
O_{11}(t,\lambda) 
= 
1 + \frac{|T_{12}|^2}{|\Lambda_1-\Lambda_2|^2} 
=
1 + \frac{|\lambda_{1,1} t_{2,12} 
+ 
t_{1,12}\lambda_{2,2}|^2}{|\lambda_{1,1}\lambda_{2,1}
-
\lambda_{1,2}\lambda_{2,2}|^2}\ .
\label{2o11}
\end{equation}
According to Eq.~\eqref{int_over_ts}, the integration over $t$'s leads to the 
following result
\begin{equation}
O_{11}(\lambda)
= 
1 
+ 
\frac{|\lambda_{1,1}|^2  
+
|\lambda_{2,2}|^2}{|\lambda_{1,1}\lambda_{2,1}
-
\lambda_{1,2}\lambda_{2,2}|^2}\ .
\end{equation}
Now we have to compute the integral over $\lambda$'s  given by Eq.~\eqref{Oz}, 
namely
\begin{equation}
O(z) 
= 
\frac{1}{2\pi^4} \int 
\left(|\lambda_{1,1}\lambda_{2,1}
-
\lambda_{1,2}\lambda_{2,2}|^2 
+
|\lambda_{1,1}|^2  
+
|\lambda_{2,2}|^2\right) 
\delta\left(z - \lambda_{1,1}\lambda_{2,1}\right) 
\prod_{i,\alpha} 
e^{-|\lambda_{i,\alpha}|^2} 
d^2 \lambda_{i,\alpha}\ .
\end{equation} 
We first integrate over the $\lambda$'s that do not appear in the Dirac delta, 
that is $\lambda_{1,2}$ and $\lambda_{2,2}$.
These integrals are in general of the Gaussian type combined with a power 
function, i.e. $\int d^2 z |z|^{2k} \exp{(-|z|^2)} = \pi k!$.
As a result of the integration, we obtain
\begin{equation}
O(z) 
= 
\frac{1}{2\pi^2} 
\int 
\left( |z|^2 + 2 + |\lambda_{1,1}|^2 \right)
\delta\left(z - \lambda_{1,1}\lambda_{2,1}\right) 
e^{-|\lambda_{1,1}|^2 -|\lambda_{2,1}|^2} 
d^2\lambda_{1,1} d^2\lambda_{2,1}\ .
\end{equation}
Now we integrate over $\lambda_{2,1}$. 
We use the scaling property of the Dirac delta
$\delta(a(z-z_0))=(1/|a|^2)\delta(z-z_0)$
to get
\begin{equation}
O(z) 
= 
\frac{1}{2\pi^2} 
\int 
\frac{|z|^2 + 2 + |\lambda_{1,1}|^2}{|\lambda_{1,1}|^2}
\exp\left(-|\lambda_{1,1}|^2-\frac{|z|^2}{|\lambda_{1,1}|^2}\right) 
d^2\lambda_{1,1}\ .
\end{equation}
The integral over $\lambda_{1,1}$ can be conveniently done in polar coordinates,
$\lambda_{1,1} = \sqrt{x} \exp{(\mathrm{i}\phi)}$
\begin{equation}
O(z) 
= 
\frac{1}{2\pi} 
\int_0^\infty 
\frac{|z|^2 + 2 + x}{x} \exp\left(-x-\frac{|z|^2}x\right) 
d x\ ,
\end{equation}
yielding
\begin{equation}
O(z) 
= 
\frac{1}{\pi} 
\left[(2 + |z|^2) K_0(2|z|) + |z| K_1(2|z|)\right]\ ,
\end{equation}
where $K_\nu$ denotes the modified Bessel function of the second kind.
The global overlap is
\begin{equation}
O 
= 
\int d^2 z\ O(z) = 2\ .
\end{equation}
The overlap density depends on the modulus $|z|$. 
It is convenient to represent this quantity as a radial function in the variable 
$r=|z|$,
\begin{equation}
O_{rad}(r) 
= 
2\pi r O(r)\ .
\label{radial_profile}
\end{equation}
Clearly $O_{rad}(r) dr$ is equal to the overlap density integrated over the 
annulus $r \le |z| \le r+dr$. 
In our case we have
\begin{equation}
O_{rad}(r) 
= 
2 r (2 + r^2) K_0(2r) + 2 r^2 K_1(2r)\ .
\label{Or22}
\end{equation}
In principle, one may repeat the calculation for any $N$ and $m$. 
All integrals except those over the $\lambda$'s appearing in the argument of 
the Dirac delta, i.e. $\delta(z-\lambda_{1,1}\cdots\lambda_{1,m})$ are 
Gaussian and can be done explicitly. 
The integrals over $\lambda$'s from the Dirac delta generate instead Meijer 
G-functions due to the multiplicative constraint~\cite{gr}.
Let us illustrate it for the product of three $2\times 2$ Ginibre matrices. 
The calculation goes as before. 
The element $T_{12}$ of the $\mathcal{T}$ matrix is
\begin{equation}
T_{12} 
= 
\lambda_{1,1} \lambda_{2,1} t_{3,12} 
+ 
\lambda_{1,1} t_{2,12} \lambda_{3,2}  
+ 
t_{1,12} \lambda_{2,2} \lambda_{3,2} \ ,
\end{equation}
and the diagonal elements are
$\Lambda_1 = \lambda_{1,1} \lambda_{2,1} \lambda_{3,1}$,
$\Lambda_2 = \lambda_{1,2} \lambda_{2,2} \lambda_{3,2}$. 
Hence, the counterpart of Eq.~\eqref{2o11} is
\begin{equation}
O_{11}(\lambda,t) 
= 
1 
+ 
\frac{|\lambda_{1,1} \lambda_{2,1} t_{3,12} 
+ 
\lambda_{1,1} t_{2,12} \lambda_{3,2}  
+ 
t_{1,12} \lambda_{2,2} \lambda_{3,2}|^2}
{|\lambda_{1,1}\lambda_{2,1}\lambda_{3,1}
-
\lambda_{1,2}\lambda_{2,2}\lambda_{3,2}|^2}\ .
\end{equation}
Integrating over $t$'s we get
\begin{equation}
O_{11}(\lambda)
= 
1 
+ 
\frac{
|\lambda_{1,1}\lambda_{2,1}|^2 
+ 
|\lambda_{1,1} \lambda_{3,2}|^2 
+ 
|\lambda_{2,2}\lambda_{3,2}|^2
}
{|\lambda_{1,1}\lambda_{2,1}\lambda_{3,1}
-
\lambda_{1,2}\lambda_{2,2}\lambda_{3,2}|^2}\ ,
\end{equation}
and over the $\lambda$'s (except those in the Dirac delta)
\begin{equation}
O(z) 
= 
\frac{1}{2\pi^6} 
\int 
\left( |z|^2 + 2 + |\lambda_{1,1} \lambda_{2,1}|^2 + |\lambda_{1,1}|^2 \right) 
\delta\left(z - \lambda_{1,1}\lambda_{2,1}\lambda_{3,1}\right) 
e^{-|\lambda_{1,1}|^2 -|\lambda_{2,1}|^2-|\lambda_{3,1}|^2} 
d^2\lambda_{1,1} d^2\lambda_{2,1} d^2\lambda_{3,1}\ .
\end{equation}
Next, we integrate over $\lambda_{3,1}$ and use polar coordinates
for $\lambda_{1,1} = \sqrt{x_1}\exp{(\mathrm{i}\phi_1)}$ and 
$\lambda_{2,1} = \sqrt{x_2} \exp{(\mathrm{i}\phi_2)}$.
We eventually obtain
\begin{equation}
O(z) 
= 
\frac{1}{2\pi} 
\int_0^\infty\int_0^\infty 
\frac{|z|^2+2+x_1x_2 + x_1}{x_1x_2}
\exp\left(-x_1 - x_2 - \frac{|z|^2}{x_1x_2} \right) 
dx_1 dx_2\ ,
\end{equation}
which yields the radial function
\begin{equation}
O_{rad}(r) 
= 
r^2 
G^{30}_{03}\left( \mbox{}^{-}_{-\frac12,\frac12,\frac12}\bigg| r^2 \right) 
+ 
2r 
G^{30}_{03}\left(\mbox{}^{-}_{0,0,0}\bigg| r^2 \right)
+ 
r 
G^{30}_{03}\left(\mbox{}^{-}_{0,0,1}\bigg| r^2 \right) 
+ 
r 
G^{30}_{03}\left(\mbox{}^{-}_{1,1,1}\bigg| r^2 \right)\ .
\label{Or23}
\end{equation}
One finds that the global overlap for $N=2$ and $m=3$ is
\begin{equation}
O 
= 
\int d^2 
z\ O(z) 
= 
\int_0^\infty O_{rad}(r) dr 
= 
\frac{5}{2}\ .
\end{equation}
One may repeat the calculations for larger $N$ and larger $m$.
The integrals one has to do are elementary but the bookkeeping gets involved and 
the calculations become tedious. 
For example, for $N=3$ and $m=2$ one has to sum three terms depending on the 
coefficients $B_1$, $B_2$ and $B_3$ as follows from Eq.~\eqref{Bs}, which 
depend on $\lambda$'s and $t$'s through  $\Lambda$'s and $T$'s:
$T_{12} = \lambda_{1,1} t_{2,12} + t_{1,12} \lambda_{2,2}$, 
$T_{13} = \lambda_{1,1} t_{2,13} + t_{1,12} t_{2,23} + t_{1,13} \lambda_{2,3}$
and $T_{23} = \lambda_{1,2} t_{2,23} + t_{1,23} \lambda_{2,3}$.
Integrals over $t$'s can be done in an algebraic way using the Wick theorem
and the following two-point functions
\begin{equation}
\langle t_{i,\alpha\beta} \bar{t}_{j,\mu\nu}\rangle_t 
= 
\delta_{ij} \delta_{\alpha\mu} \delta_{\beta\nu}  , \quad
\langle t_{i,\alpha\beta} t_{j,\mu\nu}\rangle_t 
= 
0\ ,
\end{equation}
where the symbol $\langle t_{i,\alpha\beta} \bar{t}_{j,\mu\nu}\rangle_t$ is
to be understood as follows
\begin{equation}
\langle t_{i,\alpha\beta} \bar{t}_{j,\mu\nu}\rangle_t
=
\int t_{i,\alpha\beta} \bar{t}_{j,\mu\nu}
\prod_{k,\eta<\gamma} 
\frac{1}{\pi} e^{-|t_{k,\eta\gamma}|^2} d^2 t_{k,\eta\gamma}\ .
\label{average}
\end{equation}
We skip the calculations and give the final results, which read
\begin{equation}
O_{rad}(r) 
= 
\frac{1}{3} r \left(r^4+8r^2+12\right) K_0(2r) 
+ 
\frac{1}{3} \left(2 r^4+8r^2\right) K_1(2r)
\label{Or32}
\end{equation}
and 
\begin{equation}
O 
= 
\int O_{rad}(r) dr 
=
3\ .
\label{O32}
\end{equation}
In Figs. \ref{fig:product_N2_m2}, \ref{fig:product_N2_m3} and 
\ref{fig:product_N3_m2}, we show the theoretical predictions for the radial 
profile of the overlap densities and the corresponding histograms from Monte 
Carlo simulations for $N=2,m=2$ [cf. Eq.~\eqref{Or22}], $N=2,m=3$ 
[cf. Eq.~\eqref{Or23}] and  $N=3,m=2$ [cf. Eq.~\eqref{Or32}], respectively.
We see that the Monte Carlo data follow the theoretical curves.

\section{Conjecture}

The calculations of the global density are slightly easier because there is no 
Dirac delta $\delta(z-\Lambda_1)$ in the integrand.
They are particularly simple for $N=2$.
In this case 
\begin{equation}
T_{12} 
= 
\sum_{k=1}^m t_{k,12}  
\prod_{j=1}^{k-1} \lambda_{j,1} 
\prod_{j=k+1}^{m} \lambda_{j,2}\ ,
\end{equation}
and after inserting this into Eq.~\eqref{2o11} and integrating the $t$'s, one 
obtains
\begin{equation}
O = 1 + 
\frac{1}{2\pi^{2m}} \sum_{k=1}^m \int 
\prod_{j=1}^{k-1} |\lambda_{j,1}|^2 
\prod_{j=k+1}^{m} |\lambda_{j,2}|^2  
\prod_{i=1}^m e^{-|\lambda_{i,1}|^2-|\lambda_{i,2}|^2} d^2 \lambda_{i,1} d^2 
\lambda_{i,2}\ .
\end{equation}
Each integral over $\lambda$ is either of the form 
$\int |z|^2 \exp{(-|z|^2)} d^2z=\pi$ or $\int \exp{(-|z|^2)} d^2z=\pi$, 
so all together the integration over $\lambda$'s gives the factor $\pi^{2m}$ 
which cancels the pre-factor $\pi^{-2m}$ yielding 
\begin{equation}
O 
= 
1 
+ 
\frac{m}{2}\ .
\label{O2m}
\end{equation}
Now, consider the case $m=1$ for any $N$. This case was discussed in Ref.
\cite{mc}. As follows from the discussion presented in this paper, one
can cast the overlap into the form of the following multidimensional integral
\begin{equation}
O 
= 
\frac{1}{Z} 
\int \prod_{\alpha=1}^{N-1} 
\left( 1 + \frac{1}{|\lambda_N-\lambda_\alpha|^2}\right)
|\Delta(\bm\lambda)|^2 
\prod_{\alpha=1}^N 
e^{-|\lambda_\alpha|^2} d^2 \lambda_\alpha\ ,
\label{Oandre}
\end{equation} 
where $Z=\pi^N 1!2!\cdots N!$ [cf. Eq.~\eqref{Z}]. 
What remains to do is to compute this integral.
We do this in Appendix~\ref{appendixA}, where we show that the integral 
yields
\begin{equation}
O 
= 
1 
+ 
\frac{1}{2}(N-1)\ .
\label{ON1}
\end{equation}

\begin{figure}
\includegraphics[width=15cm]{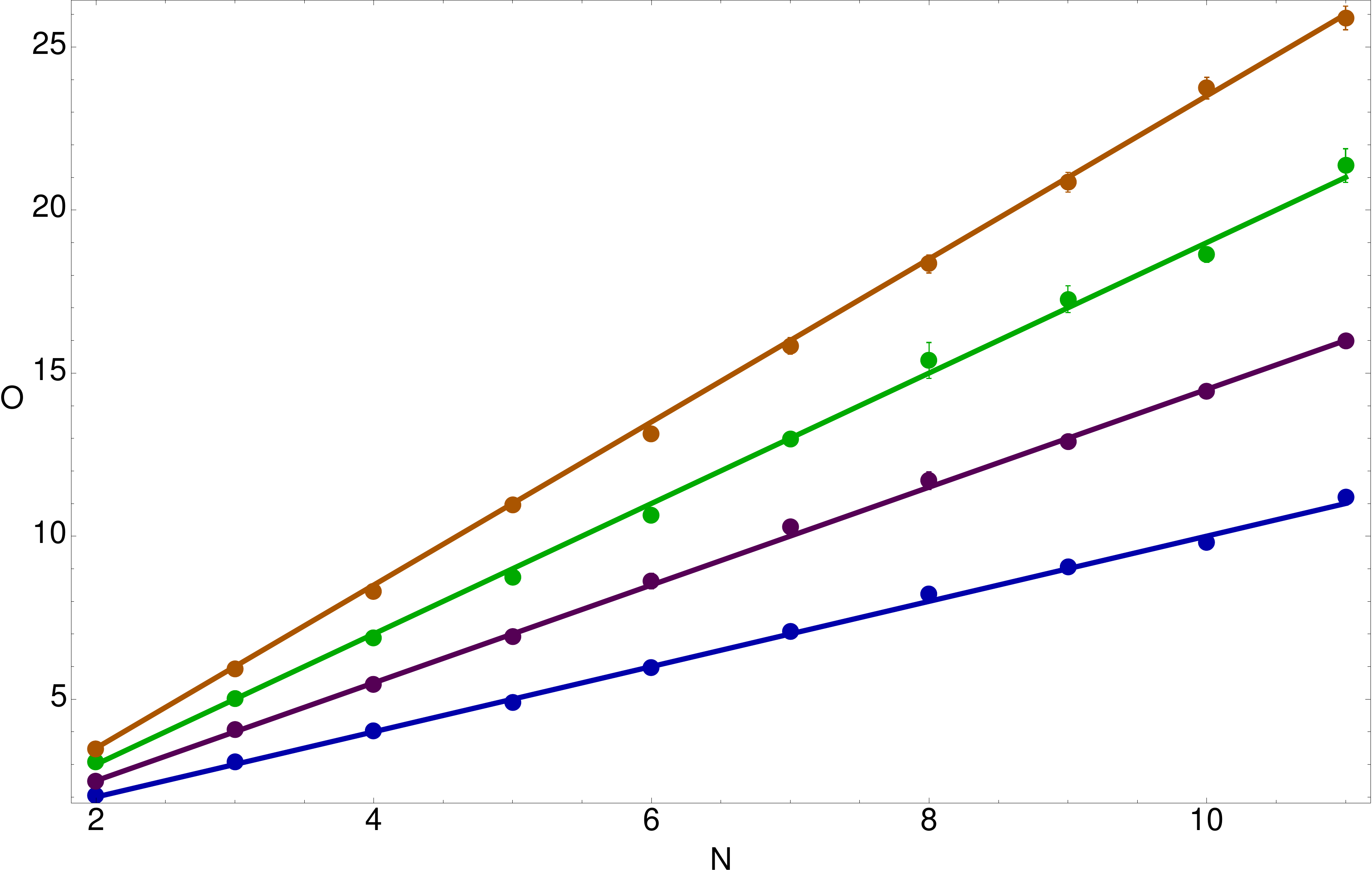}
\caption{Conjectured form of the overlap \eqref{Oconjecture} (solid lines) 
for $m=2,3,4,5$ and $N=2,\ldots, 11$ and numerical histograms (points) 
generated in Monte Carlo simulations, each for $10^4$ instances.}
\label{fig:overlap_conjecture}
\end{figure}

The results given by Eqs.~\eqref{O2m} and \eqref{ON1} suggest that $O$ grows 
linearly with $m$ and $N$, hence it is tempting to conjecture that for any $m$ 
and $N$ the global overlap is given by the formula
\begin{equation}
O 
= 
1 
+ 
\frac{m}{2}(N-1)\ . 
\label{Oconjecture}
\end{equation}
The result given by Eq.~\eqref{O32} is in agreement with this formula and Monte 
Carlo simulations fully corroborate this conjecture 
as shown in Fig.~\ref{fig:overlap_conjecture}.

\section{Large {\boldmath N} limit}

We now consider the limit $N\rightarrow \infty$. 
We set the width parameter $\sigma^2=1/N$ in the measure~\eqref{individualG}. 
The limit $N\rightarrow \infty$ has to be taken carefully since we expect
$O_N(z)$ to grow with $N$ as it results from Eq.~\eqref{Oconjecture}. 
In order to explicitly indicate the size dependence of $O(z)$ on $N$ here
we exceptionally added the subscript $N$ to $O(z)=O_N(z)$, which is implicit 
in the remaining part of the paper. It is convenient to define the 
growth rate of the overlap density as
\begin{equation}
o_N(z) 
= 
\frac{O_N(z)}{N}\ .
\end{equation}
It depends on $N$ but is expected to approach 
a $N$-independent function $o(z)$: $o_N(z) \rightarrow o(z)$ for 
$N\rightarrow \infty$. As follows from Eq.~\eqref{Oconjecture}, 
$\int d^2 z\ o(z) = m/2$. 

In the calculations, we shall use the method~\cite{bjw} that was previously
employed to calculate the limiting eigenvalue density
\begin{equation}
\rho(z) 
= 
\lim_{N\rightarrow \infty} 
\left\langle \frac{1}{N} 
\sum_{j=1}^N \delta(z-\Lambda_j) \right\rangle\ .
\end{equation}
The method is based on the generalized Green function~\cite{rj,jnpz,jnpwz}  
\begin{equation}
\label{green}
\widehat{G}(z,\epsilon) =
\left\langle 
\left(\begin{array}{cc} z \mathbbm{1}_N - X & \epsilon \mathbbm{1}_N \\
-\bar{\epsilon} \mathbbm{1}_N & \bar{z} \mathbbm{1}_N - X^\dagger 
\end{array}\right)^{-1} \right\rangle\ ,
\end{equation}
which consists of $N\times N$ blocks $G_{\alpha\beta}$
\begin{equation}
\widehat{G}(z,\epsilon) 
= 
\left(\begin{array}{cc} 
G_{11}(z,\epsilon) & G_{12}(z,\epsilon) \\ 
G_{21}(z,\epsilon) & G_{22}(z,\epsilon) \end{array} 
\right)\ .
\end{equation}
For clarity, the symbol 'hat' is reserved for matrices with a superimposed block
structure. By defining the block-trace $\mathrm{Tr_b}$ as a matrix of traces 
of individual 
blocks
\begin{equation}
\mathrm{Tr_b} \widehat{G} = 
\left(\begin{array}{cc} \mathrm{Tr}\ G_{11} & \mathrm{Tr}\ G_{12} \\ 
\mathrm{Tr}\ G_{21} & \mathrm{Tr}\ G_{22}\end{array} 
\right)\ ,
\end{equation}
one can project the $2N\times 2N$ matrix $\widehat{G}$ onto a $2\times 2$ 
matrix $\widehat{g}$
\begin{equation}
\label{gw}
\widehat{g}(z) = 
\left(\begin{array}{cc} g_{11}(z) & g_{12}(z) \\
g_{21}(z) & g_{22}(z) \end{array}\right) = 
\lim_{\epsilon \rightarrow 0} \lim_{N\rightarrow \infty}
\frac{1}{N} \mathrm{Tr_b} \widehat{G}(z,\epsilon)\ .
\end{equation}
The elements of this matrix are related to each other,
$g_{22}(z) = \bar{g}_{11}(z)$ and $g_{21}(z) = -\bar{g}_{12}(z)$ \cite{bs},  
so we have
\begin{equation}
\label{gg}
\widehat{g}(z) = \left(\begin{array}{rr} g(z) & \gamma(z) \\
-\bar{\gamma}(z) & \bar{g}(z) \end{array}\right)\ .
\end{equation}
In the large $N$ limit, the eigenvalue density is related to the diagonal element~\cite{rj,jnpz,jnpwz}  
\begin{equation}
\label{rho}
\rho(z) 
= 
\frac{1}{\pi} \frac{\partial g(z)}{\partial \bar{z}}\ ,
\end{equation}
and the growth rate of the overlap to the off-diagonal one~\cite{jnnpz}
\begin{equation}
\label{o}
o(z) 
= 
\frac{1}{\pi} \left|\gamma(z)\right|^2\ .
\end{equation}
For large $N$, the leading contribution to the overlap grows linearly 
with $N$: $O_N(z) \sim N o(z)$.

Equations~\eqref{rho} and \eqref{o} are general and can be applied to any 
random matrix provided the Green function $\widehat{g}(z)$ can be calculated. 
So the goal is now to calculate the Green function for the problem at hand. 
To this end, we use the planar diagrams enumeration 
technique~\cite{th,bipz,biz}.

\section{Dyson-Schwinger equations}
 
The enumeration of planar Feynman diagrams is a method to derive the large
$N$ limit for matrix models~\cite{th,bipz,biz}. 
The method is based on a field-theoretical representation of 
multidimensional integrals in terms of Feynman diagrams. 
One is interested in calculating the Green function 
\begin{equation}
\widehat{G}_{AB} 
= 
\left\langle\left(\widehat{Q} - \widehat{X}\right)^{-1}_{AB}
\right\rangle\ ,
\label{GAB}
\end{equation}
where $\widehat{Q}$ is a constant matrix and $\widehat{X}$ is the random matrix 
that is averaged over. 
Matrix indices are denoted by $A$ and $B$ in the last equation. 
In this approach, the Green function plays the role of generating function 
for connected two-point Feynman diagrams. 
The contributions from non-planar diagrams are suppressed at least as $1/N$ in 
the large $N$ limit, so for $N\rightarrow \infty$ only planar diagrams survive 
in the counting. 
One can write a set of equations that relate the Green function 
$\widehat{G}_{AB}$ to a generating function $\widehat{\Sigma}_{AB}$ for 
one-line irreducible diagrams.
Such equations are known in the field-theoretical literature as 
Dyson-Schwinger equations. 
Here, we are interested only in Gaussian random matrices.  
In this case, the Dyson-Schwinger equations assume a simple form
in the planar limit $N\rightarrow \infty$~\cite{bjw}
\begin{equation}
\label{DSplanar}
\begin{split}
\widehat{G}_{AB} & = \left(\widehat{Q} - \widehat{\Sigma}\right)^{-1}_{AB}\ , \\
\widehat{\Sigma}_{AD} & = \sum_{BC} \widehat{P}_{AB,CD} \widehat{G}_{BC}\ ,
\end{split}
\end{equation}
where $\widehat{P}_{AB,CD}$ represents the propagator 
\begin{equation}
\widehat{P}_{AB,CD} = \langle \widehat{X}_{AB} \widehat{X}_{CD} \rangle\ .
\end{equation}
The matrix $\widehat{Q}_{AB}$ and the propagator 
$\widehat{P}_{AB,CD}$ are inputs to be injected into these equations,
while $\widehat{G}_{AB}$ and $\widehat{\Sigma}_{AB}$ are unknown functions 
to be determined for the given inputs. 
In other words, one has first to specify what $\widehat{Q}$ and $\widehat{P}$ 
are, and then, using these equations, one can find the Green function 
$\widehat{G}$, from there $\widehat{g}$ and finally the eigenvalue density 
$\rho(z)$ [cf. Eq.~\eqref{rho}] and the overlap growth rate $o(z)$ [cf. Eq.~\eqref{o}].

\section{Single Ginibre matrix}

In this section, we review the calculations~\cite{jnpz,jnnpz} for a single 
Ginibre matrix~\cite{g}. 
In the next section, we will then show how to generalize the method to the 
product of Ginibre matrices~\cite{bjw}. 

As mentioned before, first one has to identify the matrix $\widehat{Q}$ and to 
calculate the propagator $\widehat{P}_{AB,CD}$. 
The Green function \eqref{green} reads
\begin{equation}
\widehat{G}(z,\epsilon) 
=
\left\langle  \left(\widehat{Q} - \widehat{X}\right)^{-1} \right\rangle\ ,
\end{equation}
with
\begin{equation}
\widehat{X} 
= 
\left(\begin{array}{rr} X & 0 \\
0 & X^\dagger \end{array}\right) 
\end{equation}
and
\begin{equation}
\widehat{Q} 
= 
\widehat{q} \otimes \mathbbm{1}_N\ ,
\end{equation}
where
\begin{equation}
\widehat{q} 
= 
\left(\begin{array}{rr} z & \epsilon \\
-\bar{\epsilon} & \bar{z} \end{array}\right)\ .
\end{equation}
The symbol $\otimes$ denotes the Kronecker product.
The blocks of the matrix $\widehat{X}$ can be identified with the Ginibre 
matrix and its Hermitian conjugate:
$\widehat{X}_{11} = X$, $\widehat{X}_{22} = X^\dagger$
and $\widehat{X}_{12}=\widehat{X}_{21}=0$, respectively. 
In order to calculate the propagator, we recall that the two-point 
correlations for the Ginibre matrix~\eqref{individualG} with $\sigma^2=1/N$
are 
\begin{equation}
\label{XXd}
\langle X_{ab} X^\dagger_{cd} \rangle 
= 
\int d\mu(X) X_{ab} X^{\dagger}_{cd} =
\frac{1}{N} \delta_{ad} \delta_{bc}
\end{equation}
and
\begin{equation}
\label{XX}
\langle X_{ab} X_{cd} \rangle = \langle X^\dagger_{ab} X^\dagger_{cd} 
\rangle = 0\ .
\end{equation}
Since all matrices have a block structure, it is convenient to separately write 
index positions of the blocks and positions of elements inside the blocks, and 
to split matrix indices into pairs of indices $A=(\alpha,a)$, $B=(\beta,b)$, 
$C=(\gamma,c)$, $D=(\delta,d)$, etc., with the Greek indices referring to 
the positions of the blocks, and small Latin indices to the positions within 
each block. 
The Greek indices run over the range $1$ to $2$ and the small Latin 
indices over the 
range $1$ to $N$. 
The dimension of the matrices is $2N\times 2N$. 
This block structure is also inherited by the propagators.   
Using the identification $\widehat{X}_{11} \leftrightarrow X$, 
$\widehat{X}_{22} \leftrightarrow X^\dagger$, along with Eq.~\eqref{XXd} and
\eqref{XX}, we see that the propagator factorizes into the inter-block
part (in Greek indices) and intra-block part (in Latin indices) 
\begin{equation}
\label{split}
\widehat{P}_{AB,CD} = \widehat{p}_{\alpha\beta,\gamma\delta} 
\frac{1}{N} \delta_{ad} \delta_{bc}\ .
\end{equation}
The only non-trivial elements of the inter-block part are
$\widehat{p}_{11,22}=\widehat{p}_{22,11}=1$. 
All remaining elements vanish: $\widehat{p}_{\alpha\beta,\gamma\delta}=0$. 
Since both the propagator~(\ref{split}) and the matrix 
$\widehat{Q}_{AB}=q_{\alpha\beta} \delta_{ab}$ 
are proportional to the Kronecker deltas in Latin indices, this implies that the 
matrices $\widehat{G}$ and $\widehat{\Sigma}$, being the solution of the 
Dyson-Schwinger equations~\eqref{DSplanar}, also are proportional to the 
Kronecker delta in the intra-block indices
\begin{equation}
\label{diag_solution}
\widehat{G}_{AB} = \widehat{g}_{\alpha\beta} \delta_{ab}  , \quad
\widehat{\Sigma}_{AB} = \widehat{\sigma}_{\alpha\beta} \delta_{ab}\ .
\end{equation}
Alternatively, one can write $\widehat{G} = \widehat{g} \otimes \mathbbm{1}$ 
and
$\widehat{\Sigma} =\widehat{\sigma} \otimes \mathbbm{1}$.
Therefore, one can reduce the Dyson-Schwinger  equation~\eqref{DSplanar} to 
equations for inter-block elements (in Greek indices)
\begin{equation}
\begin{split}
\left(\begin{array}{rr} {g}_{11} & 
{g}_{12} \\ {g}_{21} & {g}_{22} \end{array}\right) &=
\left(\left(\begin{array}{rr} z & \epsilon \\ 
-\bar{\epsilon}  & \bar{z}\end{array} \right) -
\left(\begin{array}{rr} {\sigma}_{11} & 
{\sigma}_{12} \\ {\sigma}_{21} & {\sigma}_{22} 
\end{array}\right)\right)^{-1} \ ,
\\
\left(\begin{array}{rr} {\sigma}_{11} & 
{\sigma}_{12} \\ {\sigma}_{21} & 
{\sigma}_{22} \end{array}\right) &=
\left(\begin{array}{cc} 0 & {g}_{12} \\ 
{g}_{21} & 0 \end{array}\right)\ .
\end{split}
\end{equation}
In the second equation, we used that 
$\widehat{p}_{11,22}=\widehat{p}_{22,11}=1$ and 
$\widehat{p}_{\alpha\beta,\gamma\delta}=0$ for other combinations of indices.
The limit $N\rightarrow \infty$ has already been taken in these equations, 
since they count contributions of planar diagrams. 
Now we can take the limit $\epsilon\rightarrow 0$ [cf. Eq.~\eqref{gw}]. 
This merely corresponds to setting $\epsilon=0$. 
Eliminating the $\{\sigma_{\alpha\beta}\}$, we get 
\begin{equation}
\label{eq_for_g}
\left(\begin{array}{cc} 
{g}_{11} & {g}_{12} \\
{g}_{21} & {g}_{22} 
\end{array}\right) = 
\left(\begin{array}{cccc} 
z & -{g}_{12}  \\
-{g}_{21} & \bar{z} \end{array}\right)^{-1}\ .
\end{equation}
Setting $g=g_{11} = \bar{g}_{22}$ and $\gamma=g_{12}=-\bar{g}_{21}$
we obtain
\begin{equation}
\label{gsolution1}
\left(\begin{array}{rr} g & \gamma \\ -\bar{\gamma} & \bar{g} \end{array}\right) = 
\left(\begin{array}{rr} z & -\gamma \\ \bar{\gamma} & \bar{z} \end{array}\right)^{-1} 
\equiv
\frac{1}{|z|^2+|\gamma|^2} 
\left(\begin{array}{rr} \bar{z} & \gamma  \\
-\bar{\gamma} & z \end{array}\right).
\end{equation}
The solution reads
\begin{equation}
\label{ggi}
g(z)=\frac{1}{z}  , \quad  \gamma(z)=0  \quad   \mathrm{for} \ |z|\ge 1
\end{equation}
and
\begin{equation}
\label{ggo}
g(z)=\bar{z}  , \quad  \left|\gamma(z)\right| = 
\sqrt{1-|z|^2} \quad \mathrm{for} \ |z| \le 1\ .
\end{equation}
The solution for $\gamma(z)$ inside the unit circle is given up to the phase, 
but this is sufficient for our purposes since the correlations density $o(z)$ 
given by Eq.~\eqref{o} depends only on the modulus of $\gamma(z)$.
Using Eqs.~\eqref{rho} and \eqref{o}, one eventually finds:
\begin{equation}
\label{rhoG1}
\rho(z) = \frac{1}{\pi} \chi_D(z)
\end{equation}
and
\begin{equation}
\label{oG1}
o(z) = \frac{1}{\pi} (1 - |z|^2) \chi_D(z)\ ,
\end{equation}
where $\chi_D$ is an indicator function for the unit disk, $\chi_D(z)=1$ for
$|z|\le 1$ and $\chi_D(z)=0$ for $|z|>1$.

\section{Product of two Ginibre matrices}
\label{sec_prod_2}

\begin{figure}
\includegraphics[width=15cm]{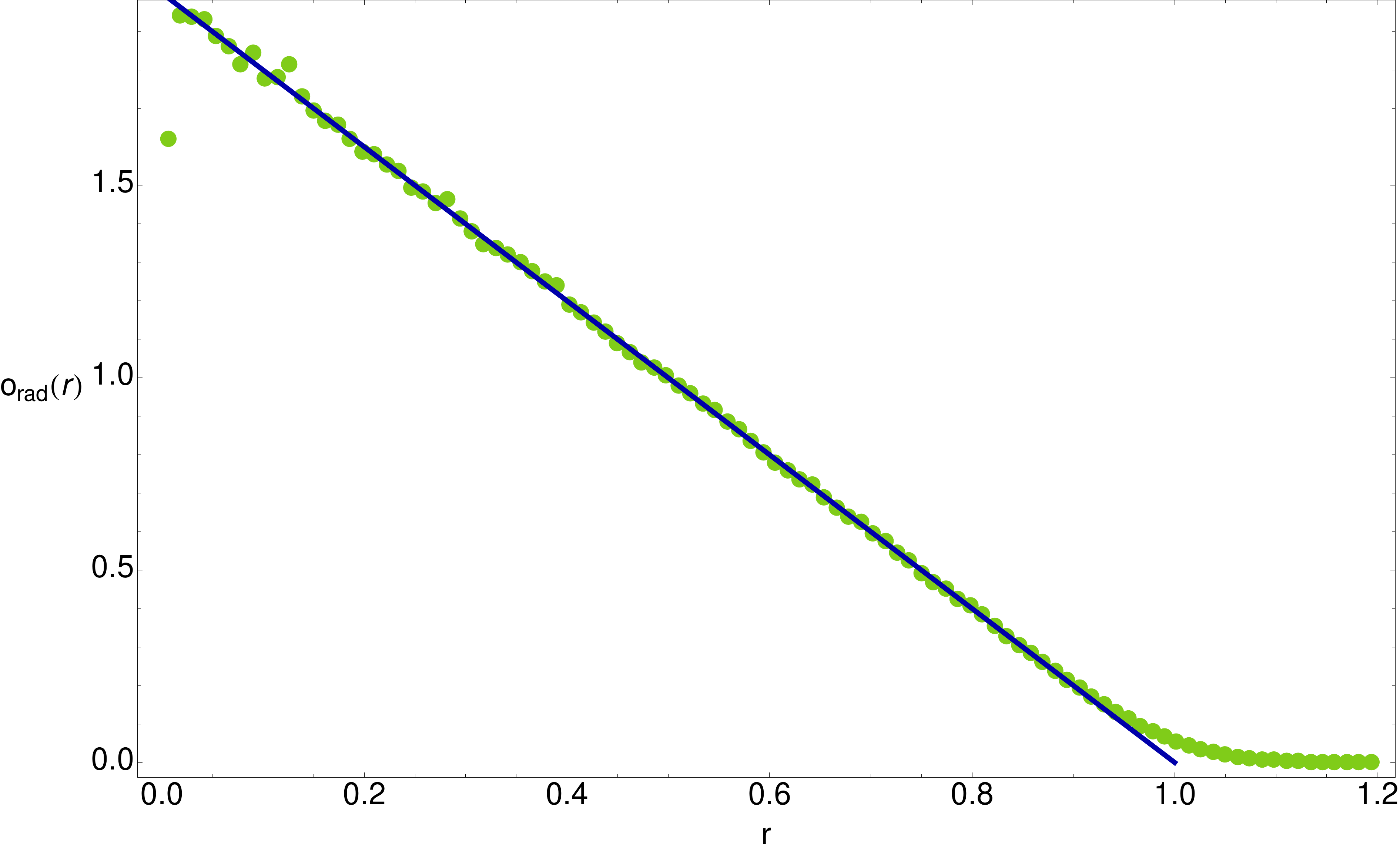}
\caption{Triangle law: theoretical prediction for  
$N\rightarrow \infty$ (\ref{orad2}) and numerical histogram (points) 
generated in Monte Carlo simulations for $10^5$ products of two
$100\times 100$ Ginibre matrices.}\label{fig:product_N100_m2}
\end{figure}

In this section, we generalize the approach from the previous section to the 
product of two Ginibre matrices~\cite{bjw}. 
The integration measure for the product $X=X_1X_2$ of independent Ginibre 
matrices $X_1$ and $X_2$ is the product of individual integration measures 
$d\mu(X_1) d\mu(X_2)$ given by Eq.~\eqref{individualG}.
According to Eq.~\eqref{XXd} the only non-vanishing two-point correlations are 
\begin{equation}
\langle X_{1,ab} X^\dagger_{1,cd} \rangle 
= 
\langle X_{2,ab} X^\dagger_{2,cd} \rangle  
= 
\frac{1}{N} \delta_{ad} \delta_{bc}\ .
\label{twoX}
\end{equation}
The Green function~\eqref{green} for the product reads
\begin{equation}
\widehat{G}(z,\epsilon) 
=
\left\langle 
\left(\begin{array}{cc} z \mathbbm{1}_N - X_1 X_2 & \epsilon \mathbbm{1}_N \\
-\bar{\epsilon} \mathbbm{1}_N & \bar{z} \mathbbm{1}_N - X_2^\dagger X_1^\dagger 
\end{array}\right)^{-1} \right\rangle\ .
\end{equation}
This form is difficult to handle because the product of Gaussian
matrices $X_1X_2$ is not Gaussian. One can however linearize the
problem by considering a block matrix of dimensions $2N \times 2N$
\begin{equation}
R 
= 
\left(\begin{array}{cc} 0 & X_1 \\ X_2 & 0 \end{array}\right)\ ,
\end{equation}
which is Gaussian. We call it root matrix because its square,
\begin{equation}
R^2 
= 
\left(\begin{array}{cc} X_1 X_2 & 0 \\ 0 & X_2 X_1 \end{array}\right) \ ,
\label{R2}
\end{equation}
reproduces two copies of the product, $X_1 X_2$ and $X_2 X_1$.
The two copies have identical eigenvalues. 
The Green function for the root matrix is
\begin{equation}
\widehat{G}(z,\epsilon) 
=
\left\langle 
\left(\begin{array}{cc} z \mathbbm{1}_N - R & \epsilon \mathbbm{1}_N \\
-\bar{\epsilon} \mathbbm{1}_N & \bar{z} \mathbbm{1}_N - R^\dagger 
\end{array}\right)^{-1} \right\rangle\ ,
\end{equation}
which is actually a $4N\times 4N$ block matrix
\begin{equation}
\widehat{G}(z,\epsilon) 
= 
\left\langle 
\left(\widehat{q} \otimes \mathbbm{1}_{N} - \widehat{R}\right)^{-1}
\right\rangle\ , 
\label{resolvent}
\end{equation}
where
\begin{equation}
\widehat{q}=\left(
\begin{array}{cccc} 
z & 0 & \epsilon & 0 \\
0 & z & 0 & \epsilon \\
-\bar{\epsilon} & 0 & \bar{z} & 0 \\
0 & -\bar{\epsilon} & 0 & \bar{z}
\end{array} \right) \stackrel{\epsilon \rightarrow 0}{\longrightarrow}
\left(\begin{array}{cccc} 
z & 0 & 0 & 0 \\
0 & z & 0 & 0 \\
0 & 0 & \bar{z} & 0 \\
0 & 0 & 0 & \bar{z}
\end{array} \right)
\end{equation}
and
\begin{equation}
\widehat{R} 
= 
\left(\begin{array}{cccc} 
0 & X_1 & 0 & 0 \\
X_2 & 0 & 0 & 0 \\
0 & 0 & 0 & X_2^\dagger \\
0 & 0 & X_1^\dagger & 0 
\end{array}\right)\ .
\label{R4}
\end{equation}
In this representation, the resolvent~\eqref{resolvent} has the standard form
in which $\widehat{R}$ is linear in the random matrices $X$'s.
Indexing blocks of $\widehat{R}$ by $\widehat{R}_{\alpha\beta}$,
with $\alpha=1,\ldots,4$ and $\beta=1,\ldots,4$, we have 
$\widehat{R}_{12}=X_1$, $\widehat{R}_{21}=X_2$,
$\widehat{R}_{34}=X_2^\dagger$, $\widehat{R}_{43}=X_1^\dagger$.
As follows from Eq.~\eqref{twoX}, the block $\widehat{R}_{12}$ is correlated 
with $\widehat{R}_{43}$ and $\widehat{R}_{21}$ with $\widehat{R}_{34}$, so the 
propagator 
\begin{equation}
\widehat{P}_{AB,CD} 
= 
\widehat{p}_{\alpha\beta,\gamma\delta} 
\frac{1}{N} \delta_{ad} \delta_{bc}
\end{equation}
has the following non-zero elements, 
$\widehat{p}_{12,43}=\widehat{p}_{43,12}=\widehat{p}_{21,34}=
\widehat{p}_{34,21}=1$. 
All other elements of $\widehat{p}_{\alpha\beta,\gamma\delta}=0$.
The situation is completely analogous to that discussed in the previous section,
except that now the problem has dimensions $4\times 4$ in inter-block indices. 
The intra-block correlations are the same as before - that is they are 
proportional to $(1/N) \delta_{ad} \delta_{bc}$ - so the solution has the
diagonal form proportional to the Kronecker delta in Latin 
indices \eqref{diag_solution}.
The Dyson-Schwinger equations~\eqref{DSplanar} for the inter-block elements
of the Green function of the root matrix read for $\epsilon\rightarrow 0$
\begin{equation}
\left(\begin{array}{cccc} 
g_{11} & g_{12} & g_{13} & g_{14} \\
g_{21} & g_{22} & g_{23} & g_{24} \\
g_{31} & g_{32} & g_{33} & g_{34} \\
g_{41} & g_{42} & g_{43} & g_{44} 
\end{array}\right) = 
\left(\begin{array}{rrrr} 
z-\sigma_{11} & -\sigma_{12} & -\sigma_{13} & -\sigma_{14} \\
-\sigma_{21} & z-\sigma_{22} & -\sigma_{23} & -\sigma_{24} \\
-\sigma_{31} & -\sigma_{32} & \bar{z}-\sigma_{33} & -\sigma_{34} \\
-\sigma_{41} & -\sigma_{42} & -\sigma_{43} & \bar{z}-\sigma_{44} 
\end{array}\right)^{-1} 
\label{ds1g}
\end{equation}
and
\begin{equation}
\left(\begin{array}{rrrr} 
\sigma_{11} & \sigma_{12} & \sigma_{13} & \sigma_{14} \\
\sigma_{21} & \sigma_{22} & \sigma_{23} & \sigma_{24} \\
\sigma_{31} & \sigma_{32} & \sigma_{33} & \sigma_{34} \\
\sigma_{41} & \sigma_{42} & \sigma_{43} & \sigma_{44} 
\end{array}\right) =
\left(\begin{array}{cccc} 
0 & 0 & g_{24} & 0 \\
0 & 0 & 0 & g_{13} \\
g_{42} & 0 & 0 & 0 \\
0 & g_{31} & 0 & 0 
\end{array}\right)\ . 
\label{ds2g}
\end{equation}
In the second equation, we used the propagator structure:
$\widehat{p}_{12,43}=\widehat{p}_{43,12}=\widehat{p}_{21,34}=
\widehat{p}_{34,21}=1$ and $\widehat{p}_{\alpha\beta,\gamma\delta}=0$
otherwise. 
Inserting $\{\sigma_{\alpha\beta}\}$ into the first equation, we get
\begin{equation}
\left(\begin{array}{cccc} 
g_{11} & g_{12} & g_{13} & g_{14} \\
g_{21} & g_{22} & g_{23} & g_{24} \\
g_{31} & g_{32} & g_{33} & g_{34} \\
g_{41} & g_{42} & g_{43} & g_{44} 
\end{array}\right) = 
\left(\begin{array}{rrrr} 
z & 0 & -g_{24} & 0 \\
0 & z & 0 & -g_{13} \\
-g_{42} & 0 & \bar{z} & 0 \\
0 & -g_{31} & 0 & \bar{z} 
\end{array}\right)^{-1} \ .
\label{gg44}
\end{equation}
It is convenient to solve this equation by defining matrices 
$\tilde{g}$ and $\tilde{\sigma}$ unitarily equivalent to $\widehat{g}$ and 
$\widehat{\sigma}$:
$\tilde{g} = P \widehat{g} P^{-1}$ and 
$\tilde{\sigma} = P \widehat{\sigma} P^{-1}$
where
\begin{equation}
P=\left(\begin{array}{cccc} 
1 & 0 & 0 & 0 \\
0 & 0 & 1 & 0 \\
0 & 1 & 0 & 0 \\
0 & 0 & 0 & 1 
\end{array}\right)\ .
\end{equation}
The effect of the similarity transformation is equivalent to permutation of 
indices of the corresponding matrices: 
$g_{\alpha\beta}=\tilde{g}_{\pi(\alpha)\pi(\beta)}$
$\sigma_{\alpha\beta}=\tilde{\sigma}_{\pi(\alpha)\pi(\beta)}$ with 
$\pi: (1,2,3,4) \rightarrow (1,3,2,4)$. 
After this transformation, Eq.~\eqref{gg44} is equivalent to
\begin{equation}
\label{44tilde}
\left(\begin{array}{cccc} 
\tilde{g}_{11} & \tilde{g}_{12} & \tilde{g}_{13} & \tilde{g}_{14} \\
\tilde{g}_{21} & \tilde{g}_{22} & \tilde{g}_{23} & \tilde{g}_{24} \\
\tilde{g}_{31} & \tilde{g}_{32} & \tilde{g}_{33} & \tilde{g}_{34} \\
\tilde{g}_{41} & \tilde{g}_{42} & \tilde{g}_{43} & \tilde{g}_{44} 
\end{array}\right) = 
\left(\begin{array}{cccc} 
z & -\tilde{g}_{34}      & 0 & 0 \\
-\tilde{g}_{43} & \bar{z} & 0 & 0  \\
0 & 0                    & z & -\tilde{g}_{12}\\
0 & 0  &  -\tilde{g}_{21}     & \bar{z} 
\end{array}\right)^{-1}\ .
\end{equation}
The matrix $\tilde{g}$ is a block matrix made of $2\times 2$ blocks. 
The off-diagonal blocks are zero while the diagonal ones fulfill the following 
equations
\begin{equation}
\left(\begin{array}{cc} 
\tilde{g}_{11} & \tilde{g}_{12} \\
\tilde{g}_{21} & \tilde{g}_{22} 
\end{array}\right) = 
\left(\begin{array}{cccc} 
z & -\tilde{g}_{34}  \\
-\tilde{g}_{43} & \bar{z} \end{array}\right)^{-1} 
\end{equation}
and 
\begin{equation}
\left(\begin{array}{cc} 
\tilde{g}_{33} & \tilde{g}_{34} \\
\tilde{g}_{43} & \tilde{g}_{44} 
\end{array}\right) = 
\left(\begin{array}{cccc} 
z & -\tilde{g}_{12}  \\
-\tilde{g}_{21} & \bar{z} \end{array}\right)^{-1}\ .
\end{equation}
The two equations admit only a symmetric solution
\begin{equation}
\left(\begin{array}{cc} 
\tilde{g}_{11} & \tilde{g}_{12} \\
\tilde{g}_{21} & \tilde{g}_{22} 
\end{array}\right) = 
\left(\begin{array}{cc} 
\tilde{g}_{33} & \tilde{g}_{34} \\
\tilde{g}_{43} & \tilde{g}_{44} 
\end{array}\right)\ ,
\end{equation}
being a solution of
\begin{equation}
\left(\begin{array}{cc} 
\tilde{g}_{11} & \tilde{g}_{12} \\
\tilde{g}_{21} & \tilde{g}_{22} 
\end{array}\right) = 
\left(\begin{array}{cccc} 
z & -\tilde{g}_{12}  \\
-\tilde{g}_{21} & \bar{z} \end{array}\right)^{-1}\ .
\end{equation}
The last equation is exactly the same as for a single Ginibre matrix 
\eqref{eq_for_g}, so the solution eventually reads
\begin{equation}
\label{tgg}
\left(\begin{array}{cccc} 
\tilde{g}_{11} & \tilde{g}_{12} & \tilde{g}_{13} & \tilde{g}_{14} \\
\tilde{g}_{21} & \tilde{g}_{22} & \tilde{g}_{23} & \tilde{g}_{24} \\
\tilde{g}_{31} & \tilde{g}_{32} & \tilde{g}_{33} & \tilde{g}_{34} \\
\tilde{g}_{41} & \tilde{g}_{42} & \tilde{g}_{43} & \tilde{g}_{44} 
\end{array}\right) = 
\left(\begin{array}{rrrr} 
g & \gamma & 0 & 0 \\
-\bar{\gamma} & \bar{g} & 0 & 0 \\
0 & 0 & g & \gamma \\
0 & 0 & -\bar{\gamma} & \bar{g} 
\end{array}\right) = \mathbbm{1}_2 \otimes 
\left(\begin{array}{rr} 
g & \gamma  \\
-\bar{\gamma} & \bar{g} 
\end{array}\right) \ ,
\end{equation}
where $g$ and $\gamma$ are given by Eqs.~\eqref{ggi} and \eqref{ggo}. 
If we permute indices back to the original order 
$\widehat{g} = P^{-1} \tilde{g} P$, 
we find 
\begin{equation}
\left(\begin{array}{cccc} 
g_{11} & g_{12} & g_{13} & g_{14} \\
g_{21} & g_{22} & g_{23} & g_{24} \\
g_{31} & g_{32} & g_{33} & g_{34} \\
g_{41} & g_{42} & g_{43} & g_{44} 
\end{array}\right) = 
\left(\begin{array}{rrrr} 
g & 0 & \gamma & 0 \\
0 & g & 0 & \gamma \\
-\bar{\gamma} & 0 & \bar{g} & 0  \\
0 & -\bar{\gamma} & 0 & \bar{g} 
\end{array}\right) = 
\left(\begin{array}{rr} 
g & \gamma  \\
-\bar{\gamma} & \bar{g} 
\end{array}\right) \otimes \mathbbm{1}_2\ .
\end{equation}
We see that the Green function for the root matrix consists of two
identical blocks equal to the Green function of a single Ginibre matrix. 
In other words, the Green function of the root matrix behaves exactly 
as a pair of copies of the Green function of a single Ginibre matrix.
The eigenvalue density and the growth rate of correlations between left and 
right eigenvectors of this matrix are given by Eqs.~\eqref{rhoG1} and 
\eqref{oG1} as
\begin{equation}
\label{rhoR}
\rho_R(z) = \frac{1}{\pi} \chi_D(z)
\end{equation}
and 
\begin{equation}
\label{oR}
o_R(z) \sim \frac{1}{\pi} (1 - |z|^2) \chi_D(z)\ .
\end{equation}
Note that the size of the root matrix is $2N\times 2N$, so the leading
term of the overlap behaves for large $N$ as 
\begin{equation}
\label{OR}
O_R(z) \sim \frac{2N}{\pi} (1 - |z|^2) \chi_D(z)\ . 
\end{equation}
From these expressions, one may derive the corresponding expressions for
$R^2$, which are directly related to the product $X_1X_2$ as follows from 
Eq.~\eqref{R2}. 
The eigenvalues of $R^2$ are related to those of $R$ as $\lambda=\lambda_R^2$,
so one can find the densities by the change of variables $z=w^2$:
$\rho(z) d^2 z = \rho_R(w) d^2 w$ 
and $O(z) d^2 z = O_R(w) d^2 w$. 
This gives
\begin{equation}
\label{rho2}
\rho(z) = \frac{1}{2\pi|z|} \chi_D(z)
\end{equation}
and
\begin{equation}
O(z) \sim \frac{N}{\pi|z|} (1-|z|) \chi_D(z)\ ,
\end{equation}
respectively. 
The result for the eigenvalue density $\rho(z)$ was first found in~\cite{bjw}. 
The overlap $O(z)$ is a new result. 
The product $X_1X_2$ is of size $N\times N$, so the growth rate is obtained
by dividing $O(z)$ by $N$,
\begin{equation}
o(z) 
= 
\lim_{N\rightarrow \infty} \frac{O(z)}{N} = 
\frac{1}{\pi|z|} \left(1-|z|\right) \chi_D(z)\ .
\end{equation}
The radial profile is obtained from the last expression by setting $r=|z|$ and 
multiplying the result by $2\pi r$ [cf. Eq.~\eqref{radial_profile}].  
This gives a triangle law
\begin{equation}
o_{rad}(r) 
= 
2 (1-r) \chi_I(r)\ ,
\label{orad2}
\end{equation}
where $\chi_I$ is an indicator function for the interval $[0,1]$:
$\chi_I(r)=1$ for $r\in [0,1]$ and $\chi_I(r)=0$ otherwise.
This prediction is compared to Monte Carlo data for $N=100$ in 
Fig.~\ref{fig:product_N100_m2}.

\begin{figure}
\includegraphics[width=15cm]{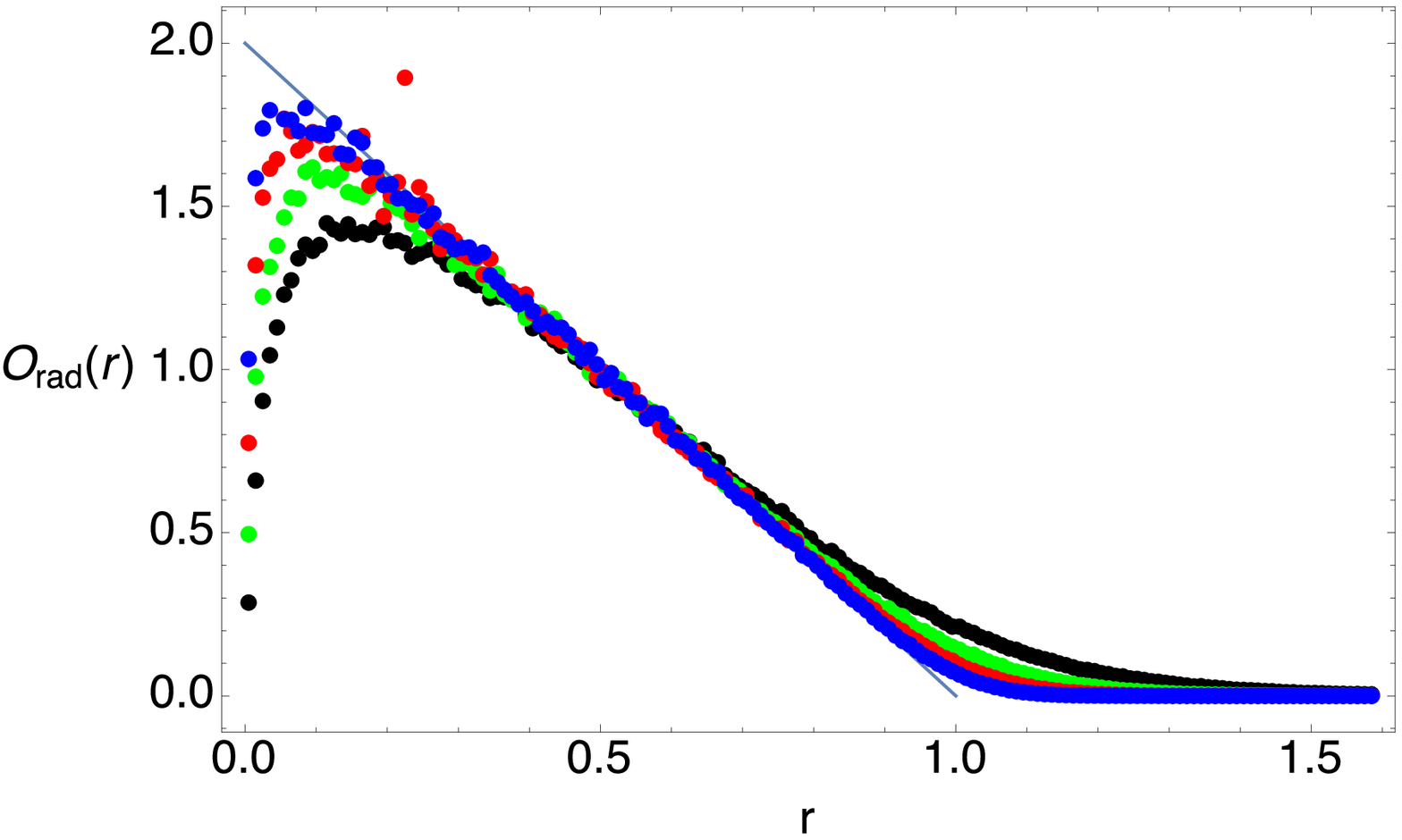}
\caption{Size dependence of the finite size corrections to the
triangle law for the product of two Ginibre matrices.
Numerical histograms are generated in Monte Carlo simulations for 
$N=10,20,40,80$ (black,green,red,blue). 
Each histogram is produced out of $2 \times 10^5$ data points.}
\label{fig:fs_effects}
\end{figure}

As one can see in the figure, there are deviations from the limiting law
for finite $N$. The radial profile drops to zero at the origin
and develops a tail going beyond the support of the limiting profile 
for large $r$. We study the $N$-dependence of these effects 
in Fig. \ref{fig:fs_effects}.
We see that the gap at the origin closes in a way characteristic of 
the hard edge behavior, while the tail at the edge of the support gets shorter 
and falls off quicker as $N$ increases. The behavior at the origin
can be probably related to the microscopic behavior of the gap probabilities, 
which are driven by the Bessel kernel and were first studied
in the context of QCD \cite{o2}. More generally, for 
the product of $m$ matrices the behavior at the origin is controlled by
the hypergeometric kernel \cite{ab}. In turn, the tail behavior at 
the soft edge is described by the error-function type 
of corrections \cite{bjw, ab}.

\section{Product of elliptic Gaussian matrices}
\label{sec:2elliptic} 

\begin{figure}
\includegraphics[width=15cm]{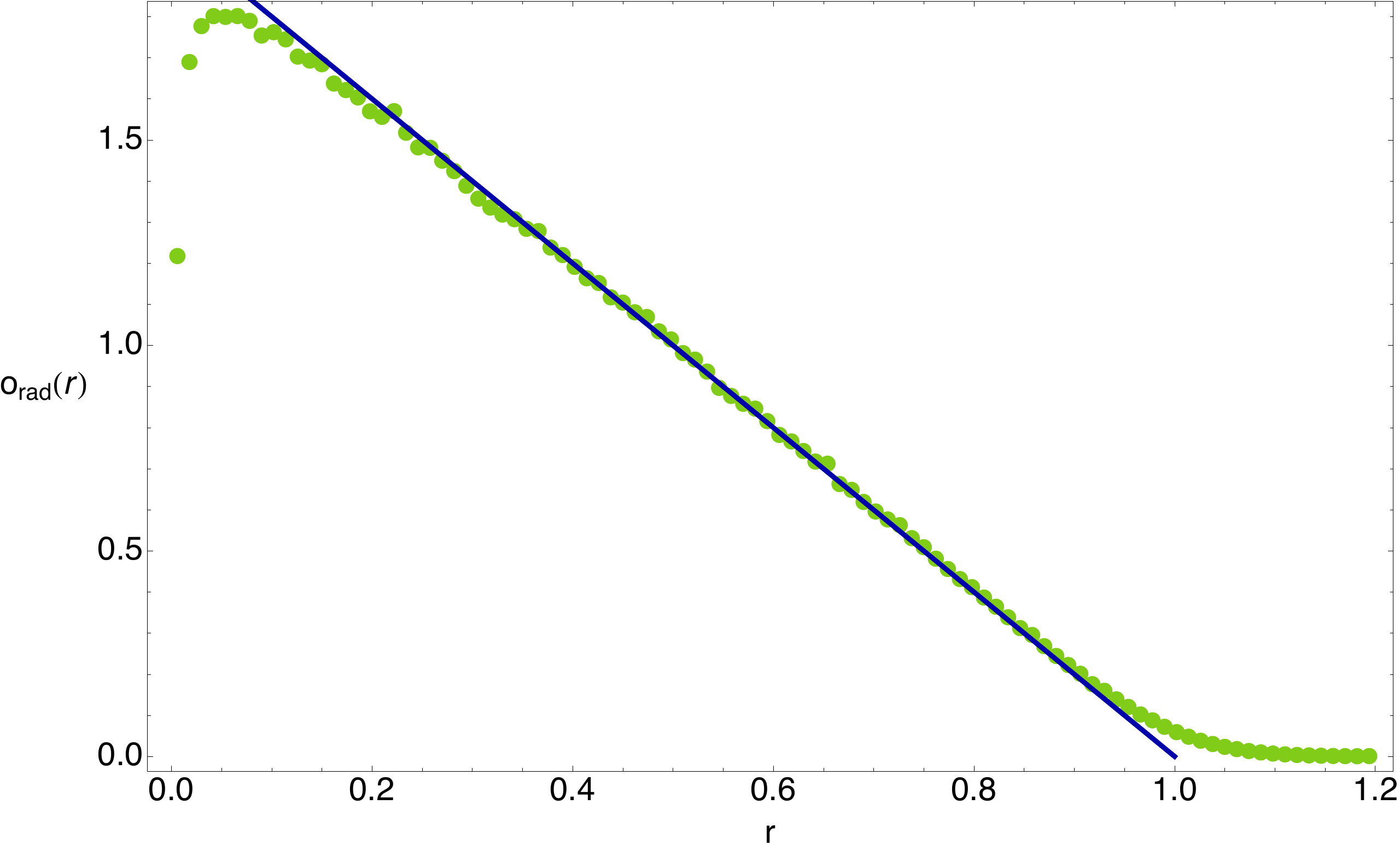}
\caption{Triangle law: theoretical prediction for  
$N\rightarrow \infty$ \eqref{orad2} and numerical histogram (points) 
generated in Monte Carlo simulations for $10^5$ products of Ginibre
times GUE matrices of dimensions $100\times 100$.}
\label{fig:product_Ginibre_GUE_N100}
\end{figure}

For completeness, we also consider the product of elliptic matrices defined by 
the measure~\cite{g1}
\begin{equation}
d\mu(X) 
= 
\frac{1}{Z} 
\exp\left[-\frac{1}{\sigma^2(1-\kappa^2)} 
\mathrm{Tr} 
\left(XX^\dagger -
\frac{\kappa}{2} (XX + X^\dagger X^\dagger)\right)\right] DX\ .
\label{elliptic}
\end{equation}
As before, we set $\sigma^2=1/N$ and scale it with $N$ while taking the limit 
$N\rightarrow \infty$.  
The parameter $\kappa$ belongs to the range $[-1,1]$. 
It is related to the ellipse eccentricity. 
For $\kappa=0$, \eqref{elliptic} reproduces the Ginibre measure. 
Generically, the support of the eigenvalue density of matrices generated 
according to the measure given by Eq.~\eqref{elliptic} is elliptic.
When $\kappa$ approaches $1$ (or $-1$), the support flattens and in the limit 
$\kappa\rightarrow \pm 1$ gets completely squeezed to an interval of the real 
(or imaginary) axis. 
The corresponding matrix becomes Hermitian (or anti-Hermitian). 
The two-point correlations for the elliptic ensemble~\eqref{elliptic} are 
\begin{equation}
\langle X_{ab} X^\dagger_{cd} \rangle = 
\langle X^\dagger_{ab} X_{cd} \rangle =
\frac{1}{N} \delta_{ad}\delta_{bc}
\label{XXed}
\end{equation}
and 
\begin{equation}
\langle X_{ab} X_{cd} \rangle 
= 
\langle X^\dagger_{ab} X^\dagger_{cd} \rangle 
= 
\kappa \frac{1}{N} \delta_{ad}\delta_{bc}\ .
\label{XXe}
\end{equation}
Consider the product $X=X_1X_2$ of two elliptic matrices $X_1$ and $X_2$
with different eccentricity parameters $\kappa_1$ and $\kappa_2$. 
As in the previous section, we construct the root matrix~\eqref{R4}, which is a 
$4N\times 4N$ matrix. 
The propagator for the root matrix elements is
\begin{equation}
\widehat{P}_{AB,CD} = \widehat{p}_{\alpha\beta,\gamma\delta} 
\frac{1}{N} \delta_{ad} \delta_{bc}\ ,
\end{equation}
where $\widehat{p}_{\alpha\beta,\gamma\delta}$ has now more nonzero elements. 
In addition to $\widehat{p}_{12,43}=\widehat{p}_{43,12}=\widehat{p}_{21,34}=
\widehat{p}_{34,21}=1$, we have
$\widehat{p}_{12,12}=\widehat{p}_{21,21}=\kappa_1$ and 
$\widehat{p}_{34,34}=\widehat{p}_{43,43}=\kappa_2$,
which come from Eq.~\eqref{XXe}.
We can now write the Dyson-Schwinger equations for this propagator. 
The first equation is identical as that for the product of Ginibre 
matrices~\eqref{ds1g}. 
The second one differs from the previous one \eqref{ds2g}, since now we have 
additional non-zero elements coming from the eccentricity parameters $\kappa_1$ and $\kappa_2$
\begin{equation}
\label{ds2e}
\left(\begin{array}{rrrr} 
\sigma_{11} & \sigma_{12} & \sigma_{13} & \sigma_{14} \\
\sigma_{21} & \sigma_{22} & \sigma_{23} & \sigma_{24} \\
\sigma_{31} & \sigma_{32} & \sigma_{33} & \sigma_{34} \\
\sigma_{41} & \sigma_{42} & \sigma_{43} & \sigma_{44} 
\end{array}\right) =
\left(\begin{array}{cccc} 
0 & \kappa_1 g_{21} & g_{24} & 0 \\
\kappa_1 g_{12} & 0 & 0 & g_{13} \\
g_{42} & 0 & 0 & \kappa_2 g_{43} \\
0 & g_{31} & \kappa_2 g_{34} & 0 
\end{array}\right)\ .
\end{equation}
Inserting the $\{\sigma_{\alpha\beta}\}$ into Eq.~\eqref{ds1g} and permuting 
indices as in the previous section, we get
\begin{equation}
\left(\begin{array}{cccc} 
\tilde{g}_{11} & \tilde{g}_{12} & \tilde{g}_{13} & \tilde{g}_{14} \\
\tilde{g}_{21} & \tilde{g}_{22} & \tilde{g}_{23} & \tilde{g}_{24} \\
\tilde{g}_{31} & \tilde{g}_{32} & \tilde{g}_{33} & \tilde{g}_{34} \\
\tilde{g}_{41} & \tilde{g}_{42} & \tilde{g}_{43} & \tilde{g}_{44} 
\end{array}\right) = 
\left(\begin{array}{cccc} 
z & -\tilde{g}_{34}      & -\kappa_1 \tilde{g}_{31} & 0 \\
-\tilde{g}_{43} & \bar{z} & 0 & -\kappa_2 \tilde{g}_{42} \\
-\kappa_1 \tilde{g}_{13} & 0                    & z & -\tilde{g}_{12}\\
0 & -\kappa_2 \tilde{g}_{24}  &  -\tilde{g}_{21}     & \bar{z} 
\end{array}\right)^{-1}\ .
\end{equation}
This equation is much more complicated than that for the product of Ginibre 
matrices~\eqref{tgg}, because the two off-diagonal blocks on the right-hand 
side are non-zero. 
However, making the {\it ansatz} that the off-diagonal blocks of the solution 
vanish 
\begin{equation}
\left(\begin{array}{cc}
\tilde{g}_{13} & \tilde{g}_{14} \\
\tilde{g}_{23} & \tilde{g}_{24} 
\end{array}\right) =
\left(\begin{array}{cc}
\tilde{g}_{31} & \tilde{g}_{32} \\ 
\tilde{g}_{41} & \tilde{g}_{42}
\end{array}\right) = 
\left(\begin{array}{cc}
0 & 0 \\
0 & 0
\end{array}\right)
\end{equation}
forces the two remaining blocks to satisfy the very same equation as for the 
product of Ginibre matrices~\eqref{44tilde},
\begin{equation}
\left(\begin{array}{cccc} 
\tilde{g}_{11} & \tilde{g}_{12} & 0 & 0 \\
\tilde{g}_{21} & \tilde{g}_{22} & 0 & 0 \\
0 & 0 & \tilde{g}_{33} & \tilde{g}_{34} \\
0 & 0 & \tilde{g}_{43} & \tilde{g}_{44} 
\end{array}\right) = 
\left(\begin{array}{cccc} 
z & -\tilde{g}_{34}      & 0 & 0 \\
-\tilde{g}_{43} & \bar{z} & 0 & 0 \\
0 & 0                    & z & -\tilde{g}_{12}\\
0 & 0  &  -\tilde{g}_{21}     & \bar{z} 
\end{array}\right)^{-1}\ ,
\end{equation}
hence the solution is the same as before.
This solution is independent of the eccentricity parameters $\kappa_1$ and 
$\kappa_2$ and moreover it is always spherically symmetric, even though the two 
matrices in the product are elliptic. 
To summarize, in the large $N$ limit the eigenvalue density and the left-right 
eigenvector correlations for the product of two elliptic matrices are 
spherically symmetric \eqref{rho2} \cite{bjw} and the eigenvector correlations 
are identical as for the product of Ginibre matrices \eqref{orad2}.
This prediction is compared to Monte Carlo data for $N=100$ in 
Fig.~\ref{fig:product_Ginibre_GUE_N100}. We see that it also follows the triangle
law as for the product of Ginibre matrices. The finite $N$ data 
exhibit however stronger finite size effects as compared to those for the product
of two Ginibre matrices which manifest as a stronger deviation from the limiting
density for small values of $r$.
Compare Figs. ~\ref{fig:product_N100_m2} and~\ref{fig:product_Ginibre_GUE_N100}.
 More generally, the limiting profile for $N\rightarrow \infty$ is independent
of $\kappa_1$ and $\kappa_2$, while the finite size corrections 
do depend on the eccentricities. We checked numerically that 
the overlap density for the product of elliptic matrices is isotropic 
(circularly invariant).

\section{Product of {\boldmath m} Ginibre matrices}

We now proceed analogously as in Sec.~\ref{sec_prod_2}, where we discussed
the product of two Ginibre matrices in the large $N$ limit. 
The integration measure for the product $X=X_1 X_2 \cdots X_m$ of $m$ 
independent Ginibre matrices $X_1, X_2, \ldots, X_m$ is the product $d\mu(X_1) 
d\mu(X_2) \cdots d\mu(X_m)$ of the individual integration 
measures given by Eq.~\eqref{individualG}.
In turn, the two-point correlations are given by Eq.~\eqref{XXd}
\begin{equation}
\label{two_point_corr}
\langle X_{\mu,ab} X^\dagger_{\nu,cd} \rangle  
= \frac{1}{N} \delta_{\mu\nu} \delta_{ad} \delta_{bc}  , \quad 
\langle X_{\mu,ab} X_{\nu,cd} \rangle = 
\langle X^\dagger_{\mu,ab} X^\dagger_{\nu,cd} \rangle = 0\ ,
\end{equation}
for $\mu,\nu = 1,\ldots,m$ and $a,b,c,d=1,\ldots,N$. 
As in Sec.~\ref{sec_prod_2}, instead of directly applying the Green 
function technique to the product $X_1X_2\cdots X_m$, we apply it to the root 
matrix $R$ being a block matrix of dimensions $mN\times mN$
\begin{equation}
\label{mRdef}
R = \left(\begin{array}{ccccc} 0 & X_1 &   0 & \ldots & 0 \\
                            0 & 0   & X_2    & \ldots & 0 \\
                              &     &        & \ddots &  \\
                            0 & 0   &  0  &  \ldots   & X_{m-1} \\
                         X_m  & 0   &  0  &  \ldots   & 0    
                            \end{array}\right)\ .
\end{equation}
The $m$-th power of the root matrix 
\begin{equation}
\label{Rm}
R^m = \left(
\begin{array}{cccc} X_1 X_2 \cdots X_m & 0 & \ldots & 0 \\
                  0 & X_2 \cdots X_m X_1 & \ldots & 0 \\
                    &                    & \ddots &  \\  
                  0 & 0                  & \ldots   & X_m X_1 \cdots X_{m-1} 
\end{array}\right)
\end{equation}
reproduces $m$ cyclic copies of the product $X_1 X_2 \cdots X_m$, which all 
have identical eigenvalues. 
The Green function for the root matrix is a $2mN\times 2mN$ block matrix
\begin{equation}
\label{m_resolvent}
\widehat{G}(z,\epsilon) = \left\langle 
\left(\widehat{q} \otimes \mathbbm{1}_{N} - \widehat{R}\right)^{-1}
\right\rangle\ , 
\end{equation}
where 
\begin{equation}
\widehat{q} = 
\left(\begin{array}{cc} 
z \mathbbm{1}_m & \epsilon \mathbbm{1}_m \\
-\bar{\epsilon} \mathbbm{1}_m & \bar{z} \mathbbm{1}_m \\
\end{array} \right) \stackrel{\epsilon\rightarrow 0}{\longrightarrow}
\left(\begin{array}{cc} 
z \mathbbm{1}_m & 0 \\
0 & \bar{z} \mathbbm{1}_m \\
\end{array} \right) 
\end{equation}
and
\begin{equation}
\label{Rmhat}
\widehat{R} = 
\left(\begin{array}{cc} 
R & 0 \\
0 & R^\dagger
\end{array}\right) =
\left(\begin{array}{cccccccccc}  0    & X_1 & 0 & \ldots & 0       & & & & &\\
                               0    & 0   & X_2 &  \ldots & 0       & & & & & \\
                                    &     &    & \ddots &  & &  & 0 & & \\               
                               0    & 0   & 0 &  \ldots & X_{m-1} & & & & & \\
                               X_m  & 0   & 0 &  \ldots & 0       & & & & & \\ 
& & & & & 0           & 0 & \ldots & 0 & X^\dagger_m               \\
& & & & & X^\dagger_1 & 0 & \ldots & 0 & 0                          \\                        
& & 0 & & & 0           & X^\dagger_2 & \ldots & 0 & 0                          \\
& & & & &             &            & \ddots & & \\  
& & & & & 0           & 0 & \ldots & X^\dagger_{m-1} & 0          
\end{array}\right)\ .
\end{equation}
The resolvent given by Eq.~\eqref{m_resolvent} has the standard form with 
$\widehat{R}$ being linear in $X$'s. 
We index blocks of $\widehat{R}$ by Greek letters $\widehat{R}_{\alpha\beta}$, 
with $\alpha,\beta=1,\ldots,2m$. 
We have the following equivalence 
$\widehat{R}_{\alpha,[\alpha]+1}\equiv X_\alpha$ 
and $\widehat{R}_{m+[\alpha]+1,m+\alpha}\equiv X_{\alpha}^\dagger$
for $\alpha=1,\ldots, m$ and $[\alpha]=\alpha$ modulo $m$. 
All other blocks are zero.
As follows from Eq.~\eqref{two_point_corr}, we see that the only non-zero 
two-point correlations are
\begin{equation}
\langle R_{\alpha,[\alpha]+1} R_{m+[\alpha]+1,m+\alpha} \rangle =
\langle X_\alpha X^\dagger_{\alpha} \rangle  , \quad 
\langle  R_{m+[\alpha]+1,m+\alpha} R_{\alpha,[\alpha]+1}\rangle =
\langle X^\dagger_{\alpha} X_\alpha \rangle\ ,
\end{equation}
for $\alpha=1,\ldots,m$. 
Thus the propagator has the form
\begin{equation}
\widehat{P}_{AB,CD} = \widehat{p}_{\alpha\beta,\gamma\delta} 
\frac{1}{N} \delta_{ad} \delta_{bc}\ ,
\end{equation}
with 
\begin{equation}
\widehat{p}_{\alpha,[\alpha]+1;m+[\alpha]+1,m+\alpha}
=
\widehat{p}_{m+[\alpha]+1,m+\alpha;\alpha,[\alpha]+1}
=
1
\end{equation}
and $\widehat{p}_{\alpha\beta,\gamma\delta}=0$ otherwise. 
The situation is analogous to that discussed in Sec.~\ref{sec_prod_2},
except that now there are $2m\times 2m$ blocks. 
The intra-block correlations are the same as before 
$(1/N) \delta_{ad} \delta_{bc}$, so the solution is given as before as 
Kronecker product with the Kronecker delta in the intra-block indices  
$\widehat{G}_{AB}=\widehat{g}_{\alpha\beta} \delta_{ab}$ 
[cf. Eq.~\eqref{diag_solution}].
The first Dyson-Schwinger equation~\eqref{DSplanar} for the inter-block 
elements of the Green function of the root matrix reads for 
$\epsilon\rightarrow 0$
\begin{equation}
\label{mds1g}
\left(\begin{array}{ccc} 
g_{1,1}   & \ldots & g_{1,2m}   \\
\vdots    & \ddots & \vdots     \\ 
g_{2m,1}  & \ldots & g_{2m,2m} 
\end{array}\right) = \left[ 
\left(\begin{array}{cc} z \mathbbm{1}_m & 0 \\ 0 & \bar{z} \mathbbm{1}_m
\end{array}\right) - 
\left(\begin{array}{ccc} 
\sigma_{1,1}  & \ldots  & \sigma_{1,2m} \\
\vdots  & \ddots  & \vdots \\
\sigma_{2m,1} & \ldots & \sigma_{2m,2m} 
\end{array}\right)\right]^{-1} \ .
\end{equation}
The second Dyson-Schwinger equation~\eqref{DSplanar} yields
\begin{equation}
\label{mds2g}
\sigma_{\alpha,m+\alpha} = g_{[\alpha]+1,m+[\alpha]+1}  , \quad
\sigma_{m+[\alpha]+1,[\alpha]+1} = g_{m+\alpha,\alpha}\ ,
\end{equation}
for $\alpha=1,\ldots,m$, and $\sigma_{\alpha\beta}=0$ for
all other elements of the matrix $\widehat{\sigma}$. 
 
\begin{figure}
\includegraphics[width=15cm]{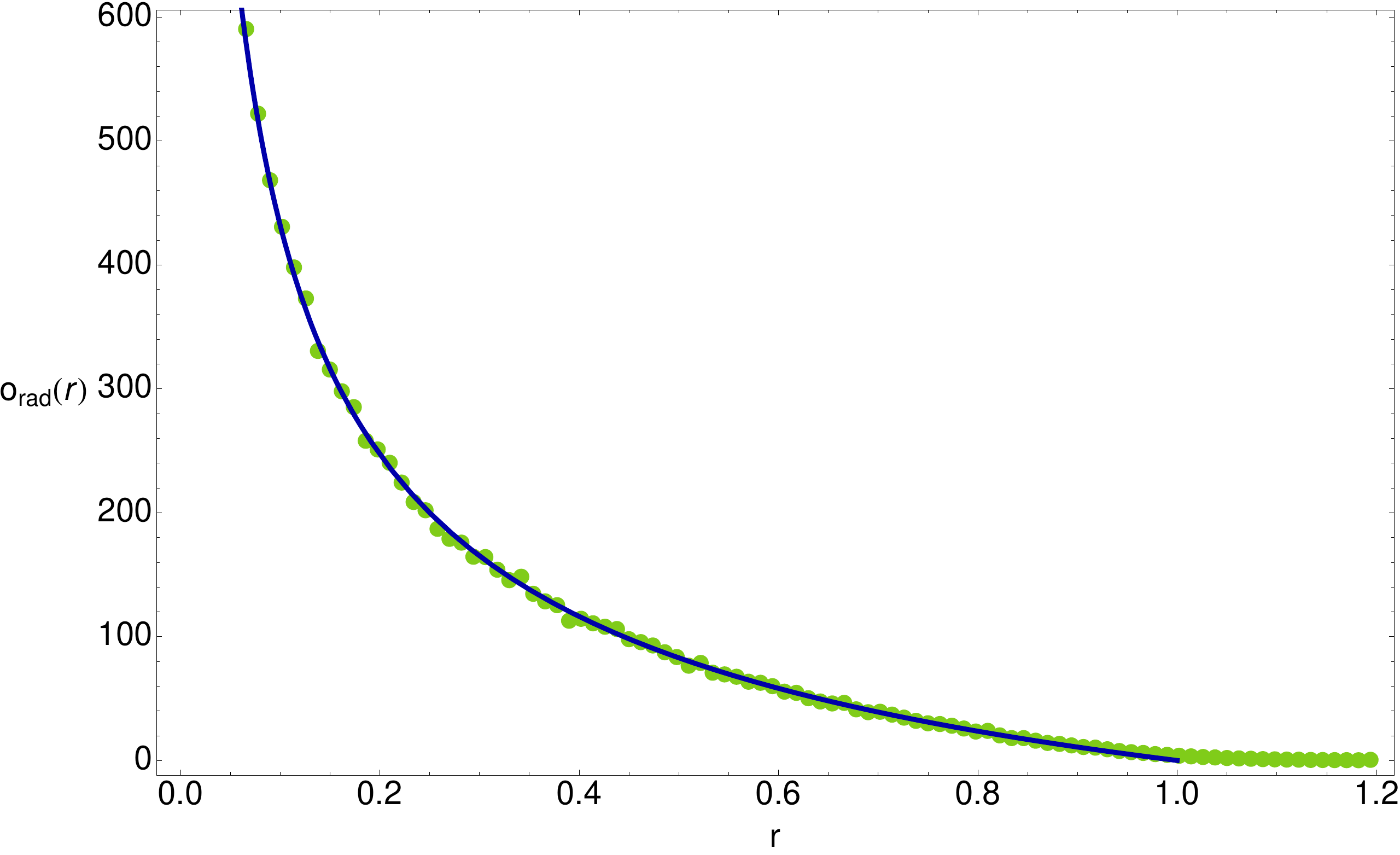}
\caption{Limiting overlap density for $m=4$: 
theoretical prediction for $N\rightarrow \infty$ \eqref{om_radial} 
and numerical histogram (points) generated in Monte Carlo simulations 
for $10^5$ products of four $100\times 100$ Ginibre matrices.}
\label{fig:product_N100_m4}
\end{figure}

The Dyson-Schwinger equations assume a simple form in a modified basis obtained 
by permutation of matrix indices, 
$\alpha \rightarrow \pi(\alpha)$, where $\pi(\alpha) = 2\alpha-1$ and 
$\pi(\alpha+m)=2\alpha$ for $\alpha=1,\ldots,m$.
We define 
$\widehat{\sigma}_{\alpha\beta} = \tilde{\sigma}_{\pi(\alpha)\pi(\beta)}$
and $\widehat{g}_{\alpha\beta} = \tilde{g}_{\pi(\alpha)\pi(\beta)}$. 
This transformation can be alternatively viewed as a similarity transformation
$\widehat{g} = P^{-1} \tilde{g} P$ and 
$\widehat{\sigma}=P^{-1} \tilde{\sigma} P$, where the elements of the matrix 
$P$ are $P_{\alpha\beta}= \delta_{\alpha\pi(\beta)}$ and 
$P^{-1}_{\alpha\beta}= \delta_{\pi(\alpha)\beta}$. 
Clearly, $\tilde{g}$ and $\widehat{g}$ as well as $\tilde{\sigma}$ and 
$\widehat{\sigma}$ are unitarily equivalent.
Equations~\eqref{mds2g} are equivalent to
\begin{equation}
\tilde{\sigma}_{2\alpha-1,2\alpha} 
= 
\tilde{g}_{(2\alpha+1),(2\alpha+2)}  , \quad
\sigma_{2\alpha,2\alpha-1} 
= 
\tilde{g}_{(2\alpha-2),(2\alpha-3)}\ ,
\end{equation}
where the function $y=(x)$ on the right hand side maps the set of integers
on the subset $\{1,2,\ldots,2m\}$ in the following way.
Any integer $x$ can be decomposed uniquely as $x=y + 2mk$ where 
$y \in \{1,2,\ldots,2m\}$ and $k$ is an integer. The function $(x)$
selects $y$ from this decomposition. In particular
$(x)=x$ and $(2m+1)=1$, $(2m+2)=2$, $(0)=2m$, $(-1)=2m-1$. 
Eliminating $\tilde{\sigma}$'s from the Dyson-Schwinger equations, we obtain
a compact equation for $\tilde{g}$'s
\begin{equation}
\label{2m2m_tilde}
\left(\begin{array}{ccccccc} 
\tilde{g}_{11} & \tilde{g}_{12} & \tilde{g}_{13} & \tilde{g}_{14} &
\ldots & \ldots & \ldots \\
\tilde{g}_{21} & \tilde{g}_{22} & \tilde{g}_{23} & \tilde{g}_{24} &
\ldots & \ldots & \ldots \\
\tilde{g}_{31} & \tilde{g}_{32} & \tilde{g}_{33} & \tilde{g}_{34} &
\ldots & \ldots & \ldots \\ 
\tilde{g}_{41} & \tilde{g}_{42} & \tilde{g}_{43} & \tilde{g}_{44} &
\ldots & \ldots & \ldots \\
\ldots & \ldots & \ldots & \ldots & \ldots & \ldots & \ldots \\
\ldots & \ldots & \ldots & \ldots & \ldots &  
\tilde{g}_{2m-1,2m-1} & \tilde{g}_{2m,2m-1} \\ 
\ldots & \ldots & \ldots & \ldots & \ldots & 
\tilde{g}_{2m-1,2m} & \tilde{g}_{2m,2m} \\
\end{array}\right) = 
\left(\begin{array}{ccccccc} 
z & -\tilde{g}_{34}      & 0 & 0 & \ldots & 0 & 0 \\
-\tilde{g}_{2m,2m-1} & \bar{z} & 0 & 0 & \ldots & 0 & 0 \\
0 & 0                    & z & -\tilde{g}_{45} & \ldots & 0 & 0\\
0 & 0  &  -\tilde{g}_{12}     & \bar{z} & \ldots & 0 & 0 \\
\ldots & \ldots & \ldots & \ldots & \ldots & \ldots & \ldots \\
0 & 0 & 0 & 0 & 0 & z & -\tilde{g}_{12}\\
0 & 0 & 0 & 0 & 0 & -\tilde{g}_{2m-2,2m-3} & \bar{z} 
\end{array}\right)^{-1} \ .
\end{equation}
The matrix $\tilde{g}$ can be viewed as a block matrix made of $2\times 2$ 
blocks. 
The off-diagonal blocks are zero and the diagonal ones fulfill the following 
equations
\begin{equation}
\left(\begin{array}{cc} 
\tilde{g}_{2\alpha-1,2\alpha-1} & \tilde{g}_{2\alpha-1,2\alpha} \\
\tilde{g}_{2\alpha,2\alpha-1} & \tilde{g}_{2\alpha,2\alpha} 
\end{array}\right) = 
\left(\begin{array}{cc} 
z & -\tilde{g}_{(2\alpha+1),(2\alpha+2)}  \\
-\tilde{g}_{(2\alpha-2),(2\alpha-3)} & \bar{z} \end{array}\right)^{-1} \ ,
\end{equation}
for $\alpha=1,\ldots,m$. 
Making the {\it ansatz} that the solution should be symmetric -  that is
$\tilde{g}_{2\alpha-1,2\alpha-1}=g$, $\tilde{g}_{2\alpha,2\alpha}=\bar{g}$,
$\tilde{g}_{2\alpha-1,2\alpha}=\gamma$ and $\tilde{g}_{2\alpha,2\alpha-1}=-\bar{\gamma}$
for all $\alpha=1,\ldots,m$, the last equations reduce to a single one
\begin{equation}
\left(\begin{array}{rr} 
g & \gamma \\ -\bar{\gamma} & \bar{g} \end{array}\right) = 
\left(\begin{array}{rr} 
z & -\gamma  \\ \bar{\gamma} & \bar{z} \end{array}\right)^{-1} \ ,
\end{equation}
which is identical as that for a single Ginibre matrix~\eqref{gsolution1}. 
Hence, the solution for $\gamma$ and $g$ is given by Eqs.~\eqref{ggi} and 
\eqref{ggo}.
This {\it ansatz} is equivalent to the one we used for $m=2$ and merely 
means that the solution should not break the symmetry between different
cyclic permutations of Ginibre matrices in the product.
Inserting the solution into $\tilde{g}$ we find
\begin{equation}
\tilde{g} = \mathbbm{1}_m \otimes 
\left(\begin{array}{rr} 
g & \gamma  \\
-\bar{\gamma} & \bar{g} 
\end{array}\right)\ ,
\end{equation}
where $g$ and $\gamma$ are given by Eqs.~\eqref{ggi} and \eqref{ggo}. 
Permuting indices back to the original order $\widehat{g} = P \tilde{g} P^{-1}$ 
\begin{equation}
\widehat{g} 
= 
\left(\begin{array}{rr} 
g & \gamma  \\
-
\bar{\gamma} & \bar{g} 
\end{array}\right) \otimes \mathbbm{1}_m\ .
\end{equation}
Hence, we see that the Green function of the root matrix behaves as $m$ 
copies of the Green function of a single Ginibre matrix.
The eigenvalue density and the growth rate of correlations between left and 
right eigenvectors of this matrix are identical as Eqs.~\eqref{rhoG1} and 
\eqref{oG1}, namely 
\begin{equation}
\rho_R(z) 
= 
\frac{1}{\pi} \chi_D(z)
\end{equation}
and
\begin{equation}
o_R(z) 
= 
\frac{1}{\pi} (1 - |z|^2) \chi_D(z)\ .
\end{equation}
The leading term of the overlap is therefore  
\begin{equation}
O_R(z) 
\sim 
\frac{mN}{\pi} (1 - |z|^2) \chi_D(z) \ .
\end{equation}
The eigenvalues $\lambda$ of $R^m$ are related to those of $R$ as 
$\lambda=\lambda_R^m$, so by changing variables as $z=w^m$ we can find
the corresponding distributions for $R^m$: $\rho(z) d^2 z = \rho_R(w) d^2 w$ 
and $O(z) d^2 z = O_R(w) d^2 w$. 
This gives
\begin{equation}
\rho(z) 
= 
\frac{1}{m\pi} |z|^{\frac{2}{m}-2} \chi_D(z)
\end{equation}
and
\begin{equation}
O(z) 
\sim 
\frac{N}{\pi} |z|^{\frac{2}{m}-2} \left(1-|z|^{\frac{2}{m}}\right) \chi_D(z)\ ,
\end{equation}
respectively. 
Thus for large $N$ the growth rate of the overlap for the product 
$X_1 X_2 \cdots X_m$ is 
\begin{equation}
o(z) 
= 
\lim_{N\rightarrow \infty} \frac{O(z)}{N} = 
\frac{1}{\pi} |z|^{\frac{2}{m}-2} \left(1-|z|^{\frac{2}{m}}\right) \chi_D(z)\ .
\end{equation}
The radial profile defined by Eq.~(\ref{radial_profile}) is
\begin{equation}
\label{om_radial}
o_{rad}(r) = 2 r^{\frac{2}{m}-1}\left(1-r^{\frac{2}{m}}\right) \chi_I(r)\ ,
\end{equation}
where as before $\chi_D$ is the indicator function for the unit disk 
$|z|\le 1$ and $\chi_I$ for the interval $[0,1]$.
While finalizing the manuscript, we learned that this result was derived 
independently in~\cite{bnst} with the aid of an extension of the Haagerup-Larsen 
theorem~\cite{fz,hl}.
The integrated growth rate is
\begin{equation}
\int o_{rad}(r) dr 
= 
\frac{m}{2}\ ,
\end{equation} 
which means that for large $N$ the overlap grows as $O \sim mN/2$ in
agreement with Eq.~\eqref{Oconjecture}.
In Fig.~\ref{fig:product_N100_m4}, we plot the expression given by 
Eq.~\eqref{om_radial} for $m=4$ and compare it to Monte Carlo data for $N=100$.  

So far we have discussed the product of $m$ Ginibre matrices.
We could repeat the whole discussion from this section for the product of 
elliptic matrices (\ref{elliptic}) with arbitrary eccentricity parameters
$\kappa_1,\kappa_2,\ldots,\kappa_m$. We would then arrive at 
an equation for $\tilde{g}$ like Eq. (\ref{2m2m_tilde}) except that the matrix on the right hand side would now have non-zero 
non-diagonal $2\times 2$ blocks. These blocks would be made of elements 
of non-diagonal blocks of $\tilde{g}$ multiplied in some way by $\kappa$'s. 
Adopting the {\rm ansatz} from Section \ref{sec:2elliptic} 
that all off-diagonal blocks of $\tilde{g}$ are equal zero we would 
reduce this equation to Eq. (\ref{2m2m_tilde}) and get the same 
result as for the product of $m$ Ginibre matrices.

\section{Conclusions}

In this paper, we have studied macroscopic and microscopic eigenvector 
statistics of the product of Ginibre matrices. 
We have developed analytical methods to calculate the left-right eigenvector 
overlap for finite $N$ and in the limit $N\rightarrow \infty$. 
The overlap is not only an interesting object from the mathematical point of 
view but is also of interest for physical problems.
In the physics literature, it is known as Petermann factor and is for example 
used as a measure of non-orthogonality of cavity modes in chaotic 
scattering~\cite{fspb,b2}. The off-diagonal overlap has been recently used 
as a sensitive indicator of non-orthogonality occurring in open systems due to perturbations resulting from shifts of resonance widths \cite{fs,g2}.   
It plays also an important role in the description of Dysonian diffusion for 
non-Hermitian random matrices~\cite{bgntw1,bgntw2}.

There are many open problems and potential generalizations of the studies
presented in this paper. 
For example, one may try to extend the studies of the microscopic eigenvector 
statistics to products of truncated unitary matrices~\cite{zs}, which can also 
be mapped onto a determinantal point process~\cite{abkn} via generalized Schur 
decomposition~\cite{ab}. 
A great challenge is to determine the microscopic eigenvalue and eigenvector 
statistics for products of elliptic matrices or to find any non-trivial 
solvable example of products of random matrices having non-spherical measures. 

We have considered complex random matrices here. It would also 
be interesting to study overlaps for products of real and quaternionic matrices.
They are much more challenging since in these cases the microscopic 
correlations are driven by Pfaffian point processes 
rather than determinantal ones. The real and quaternionic ensembles have 
additional scaling regimes near the real axis, which introduce an additional
complication. Moreover, the Schur decomposition, 
which is at the heart of the method used in this paper, cannot be applied
in a straightforward way to real matrices since 
generically they are not orthogonally similar to upper triangular ones.  
On the other hand, we believe that the limiting laws for $N\rightarrow \infty$ are identical for real and complex ensembles since the underlying
Dyson-Schwinger equations are identical in the planar limit ($N\rightarrow \infty$).

Concerning the large $N$ limit and macroscopic statistics, it would 
be interesting to generalize the calculations of the overlap to polynomials 
of random matrices~\cite{bms,bstv,bss} and to go beyond isotropic 
(R-diagonal) matrices~\cite{hl,bnst}, as well as to better understand 
the overlap in terms of the quaternionic formalism~\cite{bs}, and finally 
to calculate the off-diagonal elements of the overlap \eqref{Oo} using 
the Bethe-Salpeter equation~\cite{mc}.

\section{Acknowledgments}
We would like to thank Romuald Janik for many interesting discussions. P.V. acknowledges the stimulating research environment provided by the EPSRC
Centre for Doctoral Training in Cross-Disciplinary Approaches to Non-Equilibrium Systems (CANES, Grant No. EP/L015854/1).

\appendix
\section{Calculation of the integral \eqref{Oandre}} \label{appendixA}
In this Appendix, we detail the calculation of the integral given by 
Eq.~\eqref{Oandre}
\begin{equation}
O=
\frac{1}{Z}
\int \prod_{\alpha=1}^{N-1} 
\left(1+\frac{1}{|\lambda_N-\lambda_\alpha|^2}\right)|
\Delta_N(\bm\lambda)|^2 \prod_{\alpha=1}^N 
e^{-|\lambda_\alpha|^2}d^2\lambda_\alpha\ ,
\end{equation}
where we have renamed the Vandermonde determinant on $N$ complex variables as 
$\Delta_N(\bm\lambda)$ for convenience. 

We can rewrite this as
\begin{equation}
O
=
\frac{1}{Z}
\int \prod_{\alpha=1}^{N-1} 
\left(\frac{|\lambda_N-\lambda_\alpha|^2+1}{|\lambda_N-\lambda_\alpha|^2}\right)
|\Delta_N(\bm\lambda)|^2 \prod_{\alpha=1}^N 
e^{-|\lambda_\alpha|^2}d^2\lambda_\alpha
=
\frac{1}{Z}
\int \prod_{\alpha=1}^{N-1} 
\left(|\lambda_N-\lambda_\alpha|^2+1\right)|
\Delta_{N-1}(\bm\lambda)|^2 \prod_{\alpha=1}^N 
e^{-|\lambda_\alpha|^2}d^2\lambda_\alpha\ ,
\end{equation}
which can be more compactly expressed as
\begin{equation}
\label{OO}
O
=
\frac{1}{Z}(N-1)!
\int d^2\lambda_N e^{-|\lambda_N|^2}
\det\left(\underbrace{\int d^2 z\ e^{-|z|^2}z^{j-1}
\bar{z}^{k-1}(|\lambda_N-z|^2+1)}_{I_{jk}(\lambda_N)}\right)_{j,k=1,\ldots,N-1}\ ,
\end{equation}
using the complex version of the Andr\'eief identity~\cite{andreief}.
The integral over $z$ yields
\begin{equation}
I_{jk}(\lambda) 
=  
\pi \left((|\lambda|^2 + 1) (k-1)! + k!\right) \delta_{j,k} 
- 
\pi k!\lambda \delta_{j-1,k} 
- 
\pi (k-1)!\bar{\lambda} \delta_{j+1,k} \ .
\end{equation}
This is a tridiagonal matrix.
When calculating its determinant 
$I_{N-1}(\lambda) = \det \left(I_{jk}(\lambda)\right)_{j,k=1,\ldots,N-1}$ it is 
convenient to pull out a  common factor from each column of the matrix
\begin{equation}
I_{jk}(\lambda) 
=  
\pi (k-1)! D_{jk}(\lambda)\ ,
\end{equation}
where
\begin{equation}
D_{jk}(\lambda) 
=  
\left(|\lambda|^2 + 1 + k\right) \delta_{j,k} - 
k\lambda \delta_{j-1,k} - \bar{\lambda} \delta_{j+1,k} \ .
\end{equation}
The determinant $I_{N-1}(\lambda)$ can be related to the determinant
$D_{N-1}(\lambda) = \det\left(D_{jk}(\lambda)\right)_{j,k=1,\ldots,N-1}$ 
as follows
\begin{equation}
I_{N-1}(\lambda) 
= 
\pi^{N-1} 0!1!\cdots (N-2)! D_{N-1}(\lambda)\ .
\end{equation}
Thus we can rewrite (\ref{OO}) as  
\begin{equation}
\label{ODet}
O=\frac{1}{\pi N!}\int d^2\lambda e^{-|\lambda|^2} D_{N-1}(\lambda) \ ,
\end{equation}
where we have also replaced the normalization constant by the 
explicit expression $Z=\pi^N 1!2!\cdots N! $ [cf. Eq.~\eqref{Z}].
It remains to find the determinant $D_n(\lambda)$ for $n=N-1$.
It has the form
\begin{equation}  \label{myeq}
D_n = \left| \begin{matrix}
a_1 & b_1 & & 0\\
c_1 & \ddots & \ddots &  \\
& \ddots & \ddots &  b_{n-1} \\
0 &  & c_{n-1} & a_n  \end{matrix}  \right|\ ,
\end{equation}
with $a_n = |\lambda|^2 + 1 +n$, $b_n=-n\lambda$, $c_n=-\bar{\lambda}$.
In general the sequence $\{D_n\}$ is called \emph{continuant} and satisfies 
the following recurrence relation
$$
D_n=a_n D_{n-1}-b_{n-1}c_{n-1}D_{n-2}\ ,
$$
with initial conditions $D_0 =1$ and $D_1 =a_1$. 
In our case the recurrence takes the form
\begin{equation}
\label{recrel}
D_n
=
(|\lambda|^2 + 1 + n) D_{n-1}- (n-1) |\lambda|^2 D_{n-2}\ .
\end{equation}
The sequence $\{ D_n\}$ reveals an interesting pattern for small $n$ 
which allows us to conjecture that $D_n$ is given in closed form by
\begin{equation}
D_n(\lambda) 
= 
\sum_{k=0}^n \frac{n!(n+1-k)}{k!} |\lambda|^{2k} \ . 
\end{equation}
One can check by straightforward algebraic manipulations that
this polynomial indeed fulfills the recurrence relation \eqref{recrel}. 
The Gaussian integral of this polynomial gives a simple result
\begin{equation}
\int d^2\lambda e^{-|\lambda|^2} D_{n}(\lambda) = \pi \sum_{k=0}^n n!(n+1-k)  
= 
\pi n! \frac{(n+1)(n+2)}{2} = \pi (n+1)! \left(1+\frac{n}{2}\right) \ ,
\end{equation}
which for $n=N-1$, using \eqref{ODet}, leads to
\begin{equation}
O= 1 + \frac{1}{2}(N-1) \ ,
\end{equation}
as claimed in \eqref{ON1}.

\end{document}